\documentclass[times, review, 12pt]{elsarticle}  

\usepackage{booktabs} 
\usepackage[ruled]{algorithm2e} 
\usepackage{graphicx}
\usepackage{color, soul}  
\usepackage{rotating}
\usepackage{enumitem}
\usepackage[normalem]{ulem}
\usepackage{caption}
\usepackage[caption=false,font=footnotesize,labelformat=simple]{subfig}
\usepackage{amsmath}
\usepackage{framed}
\usepackage[table]{xcolor}
\definecolor{lightgray}{gray}{0.9}

\usepackage{pdfpages}
\usepackage{supertabular}

\usepackage{amssymb}

\journal{Information and Software Technology}

\begin{document}

\begin{frontmatter}



\title{Source Code Properties of Defective Infrastructure as Code Scripts}


\author{Akond Rahman and Laurie Williams}

\address{North Carolina State University, Raleigh, NC, USA}

\begin{abstract}
\textit{Context}: In continuous deployment, software and services are rapidly deployed to end-users using an automated deployment pipeline. Defects in infrastructure as code (IaC) scripts can hinder the reliability of the automated deployment pipeline. We hypothesize that certain properties of IaC source code such as lines of code and hard-coded strings used as configuration values, show correlation with defective IaC scripts.    \\
\textit{Objective: The objective of this paper is to help practitioners in increasing the quality of infrastructure as code (IaC) scripts through an empirical study that identifies source code properties of defective IaC scripts.} \\
\textit{Methodology}: We apply qualitative analysis on defect-related commits mined from open source software repositories to identify source code properties that correlate with defective IaC scripts. Next, we survey practitioners to assess the practitioner's agreement level with the identified properties. We also construct defect prediction models using the identified properties for 2,439 scripts collected from four datasets. \\
\textit{Results}: We identify 10 source code properties that correlate with defective IaC scripts. Of the identified 10 properties we observe lines of code and hard-coded string to show the strongest correlation with defective IaC scripts. Hard-coded string is the property of specifying configuration value as hard-coded string. According to our survey analysis, majority of the practitioners show agreement for two properties: include, the property of executing external modules or scripts, and hard-coded string. Using the identified properties, our constructed defect prediction models show a precision of 0.70$\mathtt{\sim}$0.78, and a recall of 0.54$\mathtt{\sim}$0.67.  \\
\textit{Conclusion}: Based on our findings, we recommend practitioners to allocate sufficient inspection and testing efforts on IaC scripts that include any of the identified 10 source code properties of IaC scripts.        
\end{abstract}


\end{frontmatter}


\section{Introduction}
\label{intro}

Continuous deployment is the process of rapidly deploying software or services automatically to end-users~\cite{me:agile:cd2015}. The practice of infrastructure as code (IaC) scripts is essential to implement an automated deployment pipeline, which facilitates continuous deployment~\cite{Humble:2010:CD}. Information technology (IT) organizations, such as Netflix~\footnote{https://www.netflix.com/}, Ambit Energy~\footnote{https://www.ambitenergy.com/}, and Wikimedia Commons~\footnote{https://commons.wikimedia.org/wiki/Main\_Page}, use IaC scripts to automatically manage their software dependencies, and construct automated deployment pipelines~\cite{parnin:adages}~\cite{Humble:2010:CD}. Commercial IaC tools, such as Ansible~\footnote{https://www.ansible.com/} and Puppet~\footnote{https://puppet.com/}, provide multiple utilities to construct automated deployment pipelines. Use of IaC scripts has helped IT organizations to increase their deployment frequency. For example, Ambit Energy, uses IaC scripts to increase their deployment frequency by a factor of 1,200~\footnote{https://puppet.com/resources/case-study/ambit-energy}.


Similar to software source code, the codebase for IaC scripts in an IT organization can be large, containing hundreds of lines of code~\cite{me:icst2018:iac}. IaC scripts are susceptible to human errors~\cite{parnin:adages} and bad coding practices~\cite{Cito:Cloud2015:FSE}, which can eventually introduce defects in IaC scripts~\cite{JiangAdamsMSR2015}~\cite{parnin:adages}. Defects in IaC scripts can have serious consequences for IT organizations who rely on IaC scripts to ensure reliability of the constructed automated deployment pipelines. For example in January 2017, execution of a defective IaC script erased home directories of $\mathtt{\sim}$270 users in cloud instances maintained by Wikimedia Commons~\footnote{https://wikitech.wikimedia.org/wiki/Incident\_documentation/20170118-Labs}. In our paper, we focus on identifying source code properties that correlate with defective IaC scripts. Through systematic investigation, we can identify a set of source code properties that correlate with defective scripts. Practitioners may benefit from our investigation as they can allocate sufficient inspection and testing efforts for the identified set of source code properties in IaC scripts.  

\textit{The objective of this paper is to help practitioners in increasing the quality of infrastructure as code (IaC) scripts through an empirical study that identifies source code properties of defective IaC scripts.}

We answer the following research questions: 

\begin{itemize}[leftmargin=*]
\item{\textit{\textbf{RQ-1}: What source code properties characterize defective infrastructure as code scripts?}}
\item{\textit{\textbf{RQ-2}: Do practitioners agree with the identified source code properties?}}
\item{\textit{\textbf{RQ-3}: How can we construct defect prediction models for infrastructure as code scripts using the identified source code properties?}}
\end{itemize}

We use 94 open source software (OSS) repositories and collect 12,875 commits that map to 2,439 Puppet scripts. Using 89 raters we apply qualitative analysis to determine defect-related commits. Using the defect-related commits we determine which of the 2,439 scripts are defective. We apply qualitative analysis on defect-related commits to determine which source code properties correlate with defective IaC scripts. We apply statistical analysis to empirically validate the identified properties. We conduct a survey to identify which of the identified properties practitioners agree with. Next, we build defect prediction models using the identified properties and five statistical learners: Classification and Regression Trees~\cite{cart:original}, K Nearest Neighbor classification~\cite{Tan:2005:IDM}, Logistic Regression~\cite{logit:original}, Naive Bayes classification~\cite{Tan:2005:IDM}, and Random Forest~\cite{Breiman2001:RF:ORIGINAL} to predict defective IaC scripts. We evaluate the prediction performance of the constructed prediction models using $10\times10$-fold cross validation~\cite{Tan:2005:IDM}. We also compare the prediction performance of our property-based prediction model with prediction approaches built using the bag of words technique, which is used to extract text features from IaC scripts~\cite{harris:bag:of:words:orig}.   

We list our contributions as following: 

\begin{itemize}[leftmargin=*]
\item{A ranked order of source code properties that correlate with defective IaC scripts}; 
\item{An evaluation of how practitioners perceive the identified source code properties}; and 
\item{A set of prediction models built using the identified source code properties to predict defective IaC scripts}.
\end{itemize}      

We organize the rest of the paper as following: we discuss related background and academic work in Section~\ref{bg-rel}. We discuss our methodology, datasets, and results respectively, in Section~\ref{meth}, Section~\ref{dataset}, and Section~\ref{results}. We discuss the implications of our findings in Section~\ref{discussion}. We list the limitations of our study in Section~\ref{threats}. Finally, we conclude our paper in Section~\ref{conclusion}.


\section{Background and Related Work}
\label{bg-rel}

In this section, we provide background on IaC scripts and briefly describe related academic research. 

\subsection{Background}
\label{bg}

IaC is the practice of automatically defining and managing network and system configurations, and infrastructure through source code~\cite{Humble:2010:CD}. Companies widely use commercial tools such as Puppet, to implement the practice of IaC~\cite{Humble:2010:CD}~\cite{JiangAdamsMSR2015}~\cite{ShambaughRehearsal2016}. We use Puppet scripts to construct our dataset because Puppet is considered one of the most popular tools for configuration management~\cite{JiangAdamsMSR2015}~\cite{ShambaughRehearsal2016}, and has been used by companies since 2005~\cite{propuppet:book}. Typical entities of Puppet include modules and manifests~\cite{puppet-doc}. A module is a collection of manifests. Manifests are written as scripts that use a .pp extension. 

\begin{figure}[t]
\centering
\includegraphics[scale=0.80]{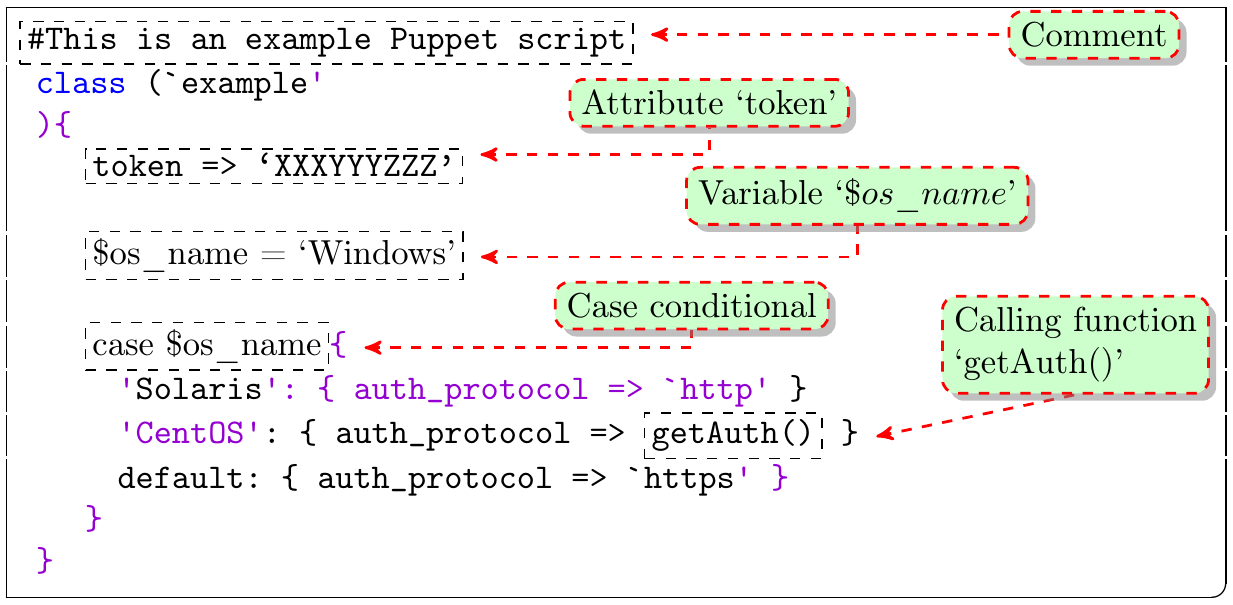}
\caption{Annotation of an example Puppet script.}
\label{fig-bg-pp}
\end{figure}

Puppet provides the utility `class' that can be used as a placeholder for the specified variables and attributes, which are used to specify configuration values. For attributes, configuration values are specified using the `$=>$' sign. For variables, configuration values are provided using the `=' sign. Similar to general purpose programming languages, code constructs such as functions/methods, comments, and conditional statements are also available for Puppet scripts. For better understanding, we provide a sample Puppet script with annotations in Figure~\ref{fig-bg-pp}.

\subsection{Related Work}
\label{rel}

Our paper is related to empirical studies that have focused on IaC technologies, such as Puppet. Sharma et al.~\cite{SharmaPuppet2016} investigated smells in IaC scripts and proposed 13 implementation and 11 design smells. Hanappi et al.~\cite{Hanappi:2016:pupp:converge} investigated how convergence of Puppet scripts can be automatically tested and proposed an automated model-based test framework. Jiang and Adams~\cite{JiangAdamsMSR2015} investigated the co-evolution of IaC scripts and other software artifacts, such as build files and source code. They reported IaC scripts to experience frequent churn. Weiss et al.~\cite{Weiss:Tortoise} proposed and evaluated `Tortoise', a tool that automatically corrects erroneous configurations in IaC scripts. Hummer at al.~\cite{Hummer:IaC} proposed a framework to enable automated testing of IaC scripts. Bent et al.~\cite{Bent:Saner2018:Puppet} proposed and validated nine metrics to detect maintainability issues in IaC scripts. Rahman et el.~\cite{Rahman:RCOSE17} investigated which factors influence usage of IaC tools. In another work, Rahman et al.~\cite{Rahman:RCOSE18} investigated the questions that programmers ask on Stack Overflow to identify the potential challenges programmers face while working with Puppet. The research study that is closest in spirit was conducted by Rahman and Williams~\cite{me:icst2018:iac}. They~\cite{me:icst2018:iac} characterized operations that appear in defective scripts, for example, setting up user account, file system operations and infrastructure provisioning. However, these identified operations do not identify which source code properties of IaC correlate with defective scripts. If we can identify source code properties that correlate with defective IaC scripts, then we can highlight source code properties which may benefit from rigorous inspection. We apply empirical analysis on identifying source code properties that correlate with defective IaC scripts. 

Our paper is also closely related to research studies that have investigated code properties that correlate with defects in source code. Nagappan and Ball~\cite{Nagappan:ICSE2005} investigated seven absolute code properties and eight relative code churn properties, and reported that relative code churn properties are better predictors of defect density. Zheng et al.~\cite{Zheng:Laurie2006} investigated how static analysis can be used to identify defects in a large scale industrial software system. They observed that the cost of automated static analysis is a relatively affordable fault detection technique, compared to that of manual inspection. Zimmermann et al.~\cite{Zimmermann:Eclipse2007} proposed a set of 14 static code properties for predicting defects in Eclipse, and reported a precision and recall of 0.63$\mathtt{\sim}$0.78, and 0.61$\mathtt{\sim}$0.78, respectively.

The above-mentioned research studies highlight the prevalence of source code properties that correlate with defects in source code. We take motivation from these studies and investigate which source code properties correlate with defective IaC scripts.


\section{Methodology}
\label{meth}

We first provide definitions, then we describe our methodology to answer our research questions.

\begin{itemize}
\item{\textbf{Defect}: An imperfection that needs to be replaced or repaired~\cite{ieee:def}.}
\item{\textbf{Defect-related commit}: A commit whose message indicates that an action was taken related to a defect.}
\item{\textbf{Defective script}: An IaC script which is listed in a defect-related commit.}
\end{itemize}


\begin{figure}[]
\centering
\subfloat[]{
  \includegraphics[width=1\linewidth]{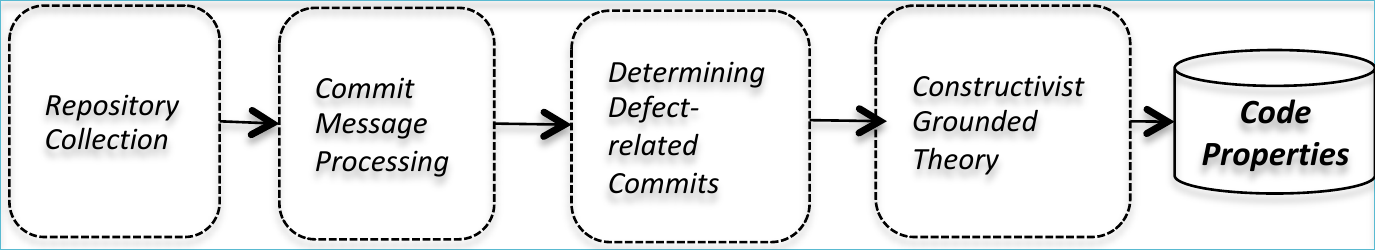}
  \label{fig:meth:rq1}
  } \\
\subfloat[]{
  \includegraphics[width=1\linewidth]{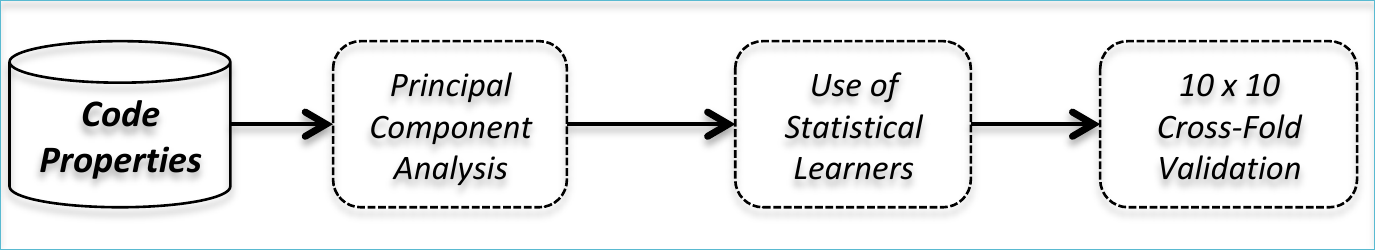}
  \label{fig:meth:rq2}
  }
\caption{Methodology. Figures~\ref{fig:meth:rq1} and~\ref{fig:meth:rq2} respectively summarizes the methodology for RQ-1 and RQ-3.}
\label{fig:meth}
\end{figure}

\subsection{\textbf{Dataset Construction}}
\label{meth-rq1}
As shown in Figure~\ref{fig:meth:rq1}, we use four steps to answer RQ-1.

\subsubsection{Repository Collection}
\label{repo-collect}


We construct IaC-specific datasets to evaluate our methodology and build prediction models. We use OSS repositories to construct our datasets by applying the following selection criteria:

\begin{itemize}[leftmargin=*]
\item{\textbf{Criteria-1}: The repository must be available for download.}
\item{\textbf{Criteria-2}: At least 11\% of the files belonging to the repository must be IaC scripts. Jiang and Adams~\citep{JiangAdamsMSR2015} reported that in OSS repositories IaC scripts co-exist with other types of files, such as Makefiles and source code files. They observed a median of 11\% of the files to be IaC scripts. By using a cutoff of 11\% we assume to collect a set of repositories that contain sufficient amount of IaC scripts for analysis.}
\item{\textbf{Criteria-3}: The repository must have at least two commits per month. Munaiah et al.~\citep{MunaiahCuration2017} used the threshold of at least two commits per month to determine which repositories have enough development activity for software organizations.}
\end{itemize} 

\subsubsection{Commit Message Processing}
\label{commit-collect}

Prior research~\cite{Ray:NaturalnessICSE2016}~\cite{Zhang:Mockus2016}~\cite{Hassan:TSE2016:AggregateMetrics} leveraged OSS repositories that use version control systems (VCS) for defect prediction studies. We use two artifacts from the VCS of the selected repositories from Section~\ref{repo-collect}, to construct our datasets: (i) commits that indicate modification of IaC scripts; and (ii) issue reports that are linked with the commits. We use commits because commits contain information on how and why a file was changed. Commits can also include links to issue reports. We use issue report summaries because they can give us more insights on why IaC scripts were changed in addition to what is found in commit messages. We collect commits and other relevant information in the following manner:

\begin{itemize}
\item{First, we extract commits that were used to modify at least one IaC script. A commit lists the changes made on one or multiple files~\cite{maletic:commit:icpc2008}.}
\item{Second, we extract the message of the commit identified from the previous step. A commit includes a message, commonly referred as a commit message. The commit messages indicate why the changes were made to the corresponding files~\cite{maletic:commit:icpc2008}.}
\item{Third, if the commit message included a unique identifier that maps the commit to an issue in the issue tracking system, we extract the identifier and use that identifier to extract the summary of the issue. We use regular expression to extract the issue identifier. We use the corresponding issue tracking API to extract the summary of the issue; and}
\item{Fourth, we combine the commit message with any existing issue summary to construct the message for analysis. We refer to the combined message as `extended commit message (XCM)' throughout the rest of the paper. We use the extracted XCMs to separate the defect-related commits from the non-defect-related commits, as described in Section~\ref{defect-related-commit}.} 
\end{itemize}
 
\subsubsection{Determining Defect-related Commits}
\label{defect-related-commit}
We use defect-related commits to identify the defective IaC scripts and the source code properties that characterizes defective IaC scripts. We apply qualitative analysis to determine which commits were defect-related commits. We perform qualitative analysis using the following three steps:

\begin{description} 
\item{\textbf{Categorization Phase}: At least two raters with software engineering experience determine which of the collected commits are defect-related. We adopt this approach to mitigate the subjectivity introduced by a single rater. Each rater determine an XCM as defect-related if it represents an imperfection in an IaC script. We provide raters with a Puppet documentation guide~\cite{puppet-doc} so that raters can obtain background on Puppet. We also provide the raters the IEEE publication on anomaly classification~\cite{ieee:def} to help raters to gather background in defect. The number of XCMs to which we observe agreements amongst the raters are recorded and the Cohen's Kappa~\cite{cohens:kappa} score is computed.}
\item{\textbf{Resolution Phase}: Raters can disagree if a commit is defect-related. In these cases, we use an additional rater's opinion to resolve such disagreements. We refer to the additional rater as the `resolver'.}
\item{\textbf{Practitioner Agreement}: To evaluate the ratings of the raters in the categorization and the resolution phase, we randomly select 50 XCMs for each dataset, and contact practitioners. We ask the practitioners if they agree to our categorization of XCMs. High agreement between the raters' categorization and programmers' feedback is an indication of how well the raters performed. The percentage of XCMs to which practitioners agreed upon is recorded and the Cohen's Kappa score is computed.} 
\end{description}

Upon completion of these three steps, we can classify which commits and XCMs are defect-related. We use the defect-related XCMs to identify the source code properties needed to answer the research questions. From the defect-related commits we determine which IaC scripts are defective, similar to prior work~\cite{Zhang:Mockus2016}. Defect-related commits list which IaC scripts were changed, and from this list we determine which IaC scripts are defective.  


\subsection{Answer to RQ-1: What source code properties characterize defective infrastructure as code scripts?}
\label{gt}

As the first step, we identify source code properties by applying qualitative analysis called constructivist grounded theory~\cite{charmaz2014:CGT}. Constructivist grounded theory is a variant of grounded theory~\cite{charmaz2014:CGT} that allows for specification of research questions, and is used to characterize properties~\cite{charmaz2014:CGT}. We use defect-related XCMs and the code changes performed in defect-related commits that we determined in Section~\ref{defect-related-commit}, to perform constructivist grounded theory. We use the defect-related XCMs because these messages can provide information on how to identify source code properties that are related to defects. We also use code changes (commonly referred to as `diffs' or `hunks') from defect-related commits because code changes report what properties of the IaC source code are changed and whether or not the changes were adding or deleting code~\cite{maletic:commit:icpc2008}.  

\begin{figure*}
\includegraphics[width=1.00\textwidth]{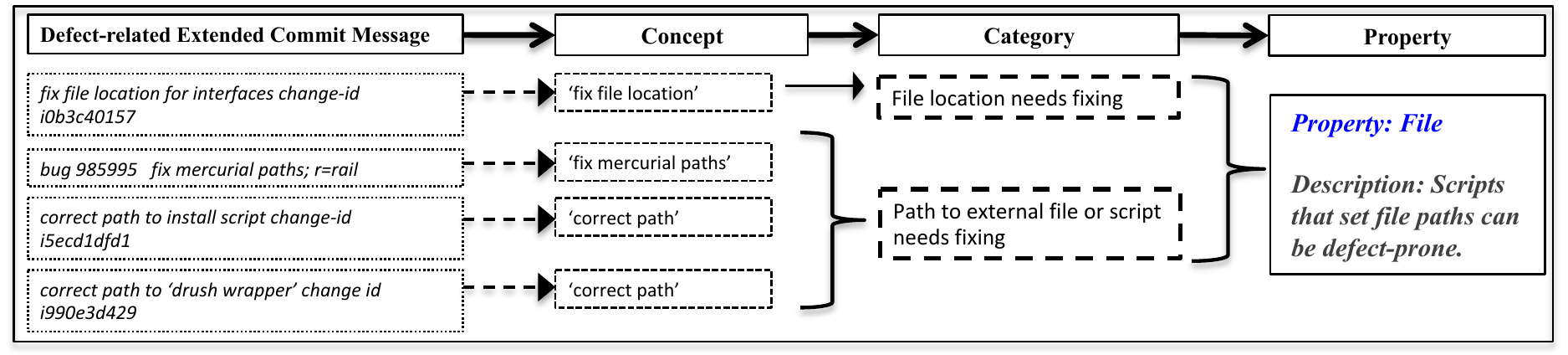}
\caption{An example of how we identify source code properties using constructivist grounded theory.}
\label{figure-cgt-res}
\end{figure*}

Any variant of grounded theory includes three elements: `concepts', `categories', and `propositions'~\cite{gt:three:elem}. In grounded theory, a proposition represents a characteristic and provides the description for the represented characteristic~\cite{gt:three:elem}. By deriving propositions, we identify properties and the description behind the identified properties. We use Figure~\ref{figure-cgt-res} to explain how we use the three grounded theory elements to identify a property. We first start with defect-related XCMs and code changes from defect-related commits, to derive concepts. According to Figure~\ref{figure-cgt-res}, from the defect-related XCM `fix file location for interfaces change-id i0b3c40157', we extract the concept `fix file location'. Next, we generate categories from the concepts, for example, we use the concept `fix file location' to determine the category which states an erroneous file location might need fixing. We use three concepts to derive category `Path to external file or script needs fixing'. Finally, we use the categories `File location needs fixing' and `Path to external file or script needs fixing' to derive a proposition related to file location. This proposition gives us a property `File' and the description behind that property is `Scripts that set file paths can be defect-prone'.  

Upon completion of constructivist grounded theory, we obtain a set of source code properties. We extract the count of each identified property using Puppeteer~\cite{SharmaPuppet2016}. We use the Mann-Whitney U test~\cite{mann:whitney:original} to compare the property count for defective and neutral files. The null hypothesis is: the property is not different between defective and neutral files. The alternative hypothesis is: property is larger for defective files than neutral files. If $p-value < 0.05$, we reject the null hypothesis, and accept the alternative hypothesis. 

Along with Mann-Whitney U test, we also apply Cliff's Delta~\cite{cliff1993:original} to compare the distribution of each characteristic between defective and neutral files. Both, Mann-Whitney U test and Cliff's Delta are non-parametric. The Mann-Whitney U test states if one distribution is significantly large/smaller than the other, whereas effect size using Cliff's Delta measures how large the difference is. 

We use Romano et al.~\cite{Romano:CliffsCutoff2006}'s recommendations to interpret the observed Cliff's Delta values. According to Romano et al.~\cite{Romano:CliffsCutoff2006}, the difference between two groups is `large' if Cliff's Delta is greater than 0.47. A Cliff's Delta value between 0.33 and 0.47 indicates a `medium' difference. A Cliff's Delta value between 0.14 and 0.33 indicates a `small' difference. Finally, a Cliff's Delta value less than 0.14 indicates a `negligible' difference.

\subsubsection{\textbf{Relative Correlation Strength of Identified Source Code Properties}}
\label{meth-fea-imp}

We use the method of `feature importance' which quantifies how important a feature is for building a prediction mode using Random Forest~\cite{Cutler:RF:FeatImp2007}. The feature importance value varies from zero to one, and a higher value for a source code property indicates higher correlation with the dependent variable. In our case the dependent variable is if a script is defective or neutral. We use Random Forests to build models using all the identified properties as independent variables, and a script of being defective or non-defective, as the dependent variable. Upon construction of the model, we compute the feature importance of each identified property provided by the Random Forest-based prediction model. To ensure stability, we follow Genuer et al.~\cite{GENUER2010:FeatImp:RF}'s recommendations and repeat the process 10 times. We report the median feature importance values for each property, and also apply the enhanced Scott-Knott test to statistically determine which property has more feature importance, and thereby exhibits more correlation with defective scripts.


\subsection{\textbf{RQ-2: Do practitioners agree with the identified source code properties?}}
\label{meth-rq2}

We conduct a survey to assess if practitioners agree with the identified set of source code properties from Section~\ref{meth-rq1}. Each of the identified properties is presented as a five-point Likert-scale question. Considering the importance of a midpoint in  Likert  scale items~\cite{themidpoint:garland}, we use a five-point scale: `Strongly Disagree', `Disagree',  `Neutral', `Agree', and `Strongly  Agree'. The survey questions are available online~\footnote{https://figshare.com/s/ad26e370c833e8aa9712}.    

While deploying, our survey we follow Internal Review Board (IRB) protocol. The IRB\# is 9379.  We deploy our survey to 350 practitioners from November 2017 to July 2018. We obtain the e-mail addresses of practitioners from the collected repositories mentioned in Section~\ref{dataset}.


\subsection{\textbf{RQ-3: How can we construct defect prediction models for infrastructure as code scripts using the identified source code properties?}}
\label{meth-rq3}

As shown in Figure~\ref{fig:meth:rq2}, in this section, we provide the methodology to answer RQ-3. We first apply log-transformation on the extracted values for each source code property. Application of log transformation on numerical features help in defect prediction, and has been used in prior research~\cite{Menzies:TSE2007}. Then, we apply principal component analysis (PCA) as described below. 

\subsubsection{Principal Component Analysis}
\label{pca}

The identified source code properties using constructivist grounded theory can show implicit correlation with each other, which needs to be accounted for. We use principal component analysis (PCA)~\cite{Tan:2005:IDM} to account for multi-collinearity amongst features~\cite{Tan:2005:IDM}, and has been extensively used in the domain of defect prediction~\cite{Nagappan:ICSE2006}~\cite{Ghotra:FeatSelect:MSR2017}. PCA creates independent linear combinations of the features that account for most of the co-variation of the features. PCA also provides a list of components and the amount of variance explained by each component. These principal components are independent and do not correlate or confound each other. We compute the total amount of variance accounted by the PCA analysis to determine what properties should be used for building prediction models. We select the principal components that account for at least 95\% of the total variance to avoid overfitting. Principal components that account for at least 95\% of the total variance, are used as input to statistical learners. 

\subsubsection{Statistical Learners}
\label{learner} 
Researchers use statistical learners to build prediction models that learn from historic data and make prediction decisions on unseen data. We use the Scikit Learn API~\cite{scikit:jml} to construct prediction models using statistical learners. We use five statistical learners that we briefly describe, and reasons for selecting these learners, as following: 

\begin{itemize}

\item{\textbf{Classification and Regression Tree (CART)}: CART generates a tree based on the impurity measure, and uses that tree to provide decisions based on input features~\cite{cart:original}. We select CART because this learner does not make any assumption on the distribution of features, and is robust to model overfitting~\cite{Tan:2005:IDM}~\cite{cart:original}. }

\item{\textbf{K Nearest Neighbor (KNN)}: The KNN classification technique stores all available prediction outcomes based on training data and classifies test data based on similarity measures. We select KNN because prior research has reported that defect prediction models that use KNN perform well~\cite{tracy:tse:lrisgood:2012}. }

\item{\textbf{Logistic Regression (LR)}: LR estimates the probability that a data point belongs to a certain class, given the values of features~\cite{logit:original}. LR provides good performance for classification if the features are roughly linear~\cite{logit:original}. We select LR because this learner performs well for classification problems~\cite{logit:original} such as defect prediction~\cite{Rahman:2013:ProcessBetter} and fault prediction~\cite{tracy:tse:lrisgood:2012}. }

\item{\textbf{Naive Bayes (NB)}: The NB classification technique computes the posterior probability of each class to make prediction decisions. We select NB because prior research has reported that defect prediction models that use NB perform well~\cite{tracy:tse:lrisgood:2012}. }

\item{\textbf{Random Forest (RF)}: RF is an ensemble technique that creates multiple classification trees, each of which are generated by taking random subsets of the training data~\cite{Breiman2001:RF:ORIGINAL}~\cite{Tan:2005:IDM}. Unlike LR, RF does not expect features to be linear for good classification performance. Researchers~\cite{Ghotra:ICSE2015} recommended the use of statistical learners that uses ensemble techniques to build defect prediction models. }

\end{itemize}

\textbf{Prediction performance measures}: We use four measures to evaluate prediction performance of the constructed models:
\begin{itemize}

\item{
\textbf{Precision}: Precision measures the proportion of IaC scripts that are actually defective given that the model predicts as defective. We use Equation~\ref{equ-pre} to calculate precision.
\begin{equation}
Precision=\frac{TP}{TP+FP}
\label{equ-pre}
\end{equation}
}
\item{
\textbf{Recall}: Recall measures the proportion of defective IaC scripts that are correctly predicted by the prediction model. We use Equation~\ref{equ-rec} to calculate recall.
\begin{equation}
Recall=\frac{TP}{TP+FN}
\label{equ-rec}
\end{equation}
}
\item{\textbf{Area Under The Receiver Operating Characteristic Curve (AUC)}: AUC uses the receiver operating characteristic (ROC). ROC  is a two-dimensional curve that plots the true positive rates against false positive rates. An ideal prediction model's ROC curve has an area of 1.0. A random prediction's ROC curve has an area of 0.5. We refer to the \textit{a}rea \textit{u}nder the ROC \textit{c}urve as AUC throughout the paper. We consider AUC as this measure is threshold independent unlike precision and recall~\cite{Ghotra:ICSE2015}, and recommended by prior research~\cite{Lessmann:TSE2008}.} 

\item{\textbf{F-Measure}: F-Measure is the harmonic mean of precision and recall. Increase in precision, often decreases recall, and vice-versa~\cite{Menzies:TSE2007:PRECISION:INSTABILITY}. F-Measure provides a composite score of precision and recall, and is high when both precision and recall is high.} 

\end{itemize}  

\textbf{Comparing prediction performance}:  To determine if parameter tuning statistically improves performance we use a variant of the \textit{S}cott \textit{K}nott (SK) test~\cite{Chakkrit:TSE2017}. This variant of SK does not assume input to be normal, and accounts for negligible effect size~\cite{Chakkrit:TSE2017}. SK uses hierarchical clustering analysis to partition the input data into significantly ($\alpha=0.05$) distinct ranks~\cite{Chakkrit:TSE2017}. According to SK, a learner for which parameter tuning is applied ranks higher if prediction performance is significantly higher. For example, if tuned CART ranks higher than that of non-tuned CART, then we can state that parameter tuning significantly increases prediction performance. We use SK to compare if parameter tuning significantly increases AUC and F-Measure for both statistical learners, and for the evaluation methods used.

\subsubsection{Evaluation Methods}
\label{eval}

We use 10$\times$10-fold cross validation to evaluate our prediction models. We use the 10$\times$10-fold cross validation evaluation approach by randomly partitioning the dataset into 10 equal sized subsamples or folds~\cite{Tan:2005:IDM}. The performance of the constructed prediction models are tested by using nine of the 10 folds as training data, and the remaining fold as test data. Similar to prior research~\cite{Ghotra:ICSE2015}, we repeat the 10-fold cross validation 10 times to assess the statistical learner's prediction stability. We report the median prediction performance score of the 10 runs.

\subsubsection{Comparison Model Construction} 
\label{baseline} 

As a comparison, we use text feature-based technique. In recent work, Rahman and Williams~\cite{me:icst2018:iac} have reported that certain text features can be used to characterize defective IaC scripts, and to build models to predict defective IaC scripts. Their findings are consistent with prior work in other domains, which has shown that text features are correlated with defects, and good predictors of defective artifacts~\cite{walden:issre:vpm:tm}. We use the `bag-of-words (BOW)'~\cite{harris:bag:of:words:orig} technique to construct prediction models to compare our identified property-based prediction models. The BOW technique which has been extensively used in software engineering~\cite{walden:issre:vpm:tm}, converts each IaC script in the dataset to a set of words or tokens, along with their frequencies. Using the frequencies of the collected tokens we create features. 

\textbf{Text Pre-processing:} Before creating text features using bag-of-words, we apply the following text pre-processing steps: 
\begin{itemize}[leftmargin=*]
\item{First, we remove comments from scripts. }
\item{Second, we split the extracted tokens according to naming conventions: camel case, pascal case, and underscore. These splitted tokens might include numeric literals and symbols, so we remove these numeric literals and symbols. We also remove stop words.}
\item{Finally, we apply Porter stemming~\cite{porter:original:1997} on the collected tokens. After completing the text pre-processing step we collect a set of pre-processed tokens for each IaC script in each dataset. We use these sets of tokens to create feature vectors as shown in Section~\ref{meth:bow}.}
\end{itemize}    

\textbf{Bag-of-Words (BOW)}:
\label{meth:bow}
Using the BOW technique, we use the tokens extracted from text pre-processing step. We compute the occurrences of tokens for each script. By using the occurrences of tokens we construct a feature vector. Finally, for all the scripts in the dataset we construct a feature matrix.

\begin{figure}[htbp]
\centering
\includegraphics[scale=0.95]{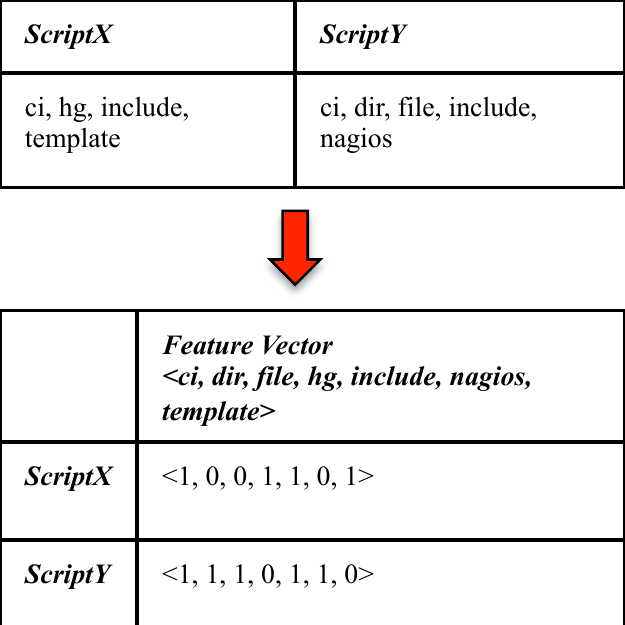}
\caption{A hypothetical example to illustrate the BOW technique discussed in Section~\ref{meth:bow}.}
\label{fig:meth:bag:of:words}
\end{figure}

We use a hypothetical example shown in Figure~\ref{fig:meth:bag:of:words} to illustrate the BOW technique. In our hypothetical example, our dataset has two IaC scripts $ScriptX$ and $ScriptY$ that respectively contain four and five pre-processed tokens. From the occurrences of tokens, we construct a feature matrix where the the token `ci' appears once for $ScriptX$ and $ScriptY$. 

Similar to our properties derived from constructivist grounded theory, we construct defect prediction models using CART, LR, NB, and RF. We compute prediction performance using 10$\times$10-fold cross validation, and compute precision, recall, AUC, and F-Measure. We use the Enhanced Scott-Knott test to compare if the our properties derived using the constructivist grounded theory process, significantly outperforms the text feature-based analysis. If the text feature-based analysis is better than our derived properties, then the Scott-Knott test will rank the text feature-based analysis higher.


\section{Datasets}
\label{dataset}

We construct datasets using Puppet scripts from OSS repositories maintained by four organizations: Mirantis, Mozilla, Openstack, and Wikimedia Commons. We select Puppet because it is considered as one of the most popular tools to implement IaC~\cite{JiangAdamsMSR2015}~\cite{ShambaughRehearsal2016}, and has been used by organizations since 2005~\cite{propuppet:book}. Mirantis is an organization that focuses on the development and support of cloud services such as OpenStack~\footnote{https://www.mirantis.com/}. Mozilla is an OSS community that develops, uses, and supports Mozilla products such as Mozilla Firefox~\footnote{https://www.mozilla.org/en-US/}. Openstack foundation is an open-source software platform for cloud computing where virtual servers and other resources are made available to customers~\footnote{https://www.openstack.org/}. Wikimedia Foundation is a non-profit organization that develops and distributes free educational content~\footnote{https://wikimediafoundation.org/}. 

\subsection{Repository Collection}

We apply the three selection criteria presented in Section~\ref{repo-collect} to identify the repositories that we use for analysis. We describe how many of the repositories satisfied each of the three criteria in Table~\ref{table-criteria-dataset}. Each row corresponds to the count of repositories that satisfy each criteria. For example, 26 repositories satisfy Criteria-1, for Mirantis. We obtain 94 repositories to extract Puppet scripts from. 

\begin{table}[]
\centering
\caption{Filtering Criteria to Construct Defect Datasets}
\label{table-criteria-dataset}
\footnotesize
{
\begin{tabular}{ p{2.0cm}  p{2.5cm} p{2.5cm} p{2.5cm} p{2.5cm} }
\hline
\textbf{Criteria}   & \textbf{Dataset} \\
\hline
                    & Mirantis & Mozilla & Openstack & Wikimedia \\
\hline
\textbf{Criteria-1} & 26 & 1,594 & 1,253 & 1,638 \\
\textbf{Criteria-2} & 20 & 2     & 61 & 11 \\
\textbf{Criteria-3} & 20 & 2     & 61 & 11 \\
\hline
\textbf{Final}      & 20 & 2     & 61 & 11 \\
\hline
\end{tabular}
}
\end{table}

\subsection{Commit Message Processing}

We report summary statistics on the collected repositories in Table~\ref{table-defect-dataset}. According to Table~\ref{table-defect-dataset}, for Mirantis we collect 165 Puppet scripts that map to 1,021 commits. Of these 1,021 commits, 82 commits include identifiers for bug reports. The constructed datasets used for empirical analysis are available as online~\footnote{https://figshare.com/s/ad26e370c833e8aa9712}.   


\begin{table*}[]
\centering
\caption{Summary statistics of constructed datasets}
\label{table-defect-dataset}
{\footnotesize
\begin{tabular}{ p{3.0cm} p{2.0cm} p{2.0cm} p{2.0cm} p{2.0cm} }
\hline
\textbf{Statistic}  & \textbf{Dataset} \\
                     & Mirantis & Mozilla & Openstack & Wikimedia \\
\hline
\textbf{Puppet Scripts} & 180 & 580  & 1,383 & 296 \\
\hline
\textbf{Defective Puppet Scripts} & 91 of 180, 50.5\% & 259 of 580, 44.6\%  & 810 of 1383, 58.5\% & 161 of 296, 54.4\% \\
\hline
\textbf{Commits with Puppet Scripts} & 1,021  & 3,074  & 7,808  & 972  \\
\hline
\textbf{Commits with Report IDs} & 82  of 1021, 8.0\% & 2764 of 3074, 89.9\% & 2252 of 7808, 28.8\% & 210 of 972, 21.6\% \\
\hline
\textbf{Defect-related Commits} &  344 of 1021, 33.7\% & 558 of 3074, 18.1\%  & 1987 of 7808, 25.4\% & 298 of 972, 30.6\% \\
\hline
\end{tabular}
}
\end{table*}

\subsection{Determining Categories of Defects}
\label{res-study-dataset}
We use 89 raters to categorize the XCMs, using the following phases:

\begin{itemize}
\item{
\textbf{Categorization Phase}: 
\begin{itemize} 
\item{\textbf{Mirantis}: We recruit students in a graduate course related to software engineering via e-mail. The number of students in the class was 58, and 32 students agreed to participate. We follow IRB\#12130, in recruitment of students and assignment of defect categorization tasks. We randomly distribute the 1,021 XCMs amongst the students such that each XCM is rated by at least two students. The average professional experience of the 32 students in software engineering is 1.9 years. On average, each student took 2.1 hours.}
\item{\textbf{Mozilla}: One second year PhD student and one fourth year PhD student separately apply qualitative analysis on 3,074 XCMs. The fourth and second year PhD student, respectively, have a professional experience of three and two years in software engineering. The fourth and second year PhD student, respectively, took 37.0 and 51.2 hours to complete the categorization. }
\item{\textbf{Openstack}: One second year PhD student and one first year PhD student separately, apply qualitative analysis on 7,808 XCMs from Openstack repositories.The second and first year PhD student respectively, have a professional experience of two and one years in software engineering. The second and first year PhD student completed the categorization of the 7,808 XCMs respectively, in 80.0 and 130.0 hours.}
\item{\textbf{Wikimedia}: 54 graduate students recruited from the `Software Security' course are the raters. We randomly distribute the 972 XCMs amongst the students such that each XCM is rated by at least two students. According to our distribution, 140 XCMs are assigned to each student. The average professional experience of the 54 students in software engineering is 2.3 years. On average, each student took 2.1 hours to categorize the 140 XCMs. The IRB protocol was IRB\#9521. }
\end{itemize}
}
\item{
\textbf{Resolution Phase}: 
\begin{itemize} 
\item{\textbf{Mirantis}: Of the 1,021 XCMs, we observe agreement for 509 XCMs and disagreement for 512 XCMs, with a Cohen's Kappa score of 0.21. Based on Cohen's Kappa score, the agreement level is `fair'~\citep{Landis:Koch:Kappa:Range}.}
\item{\textbf{Mozilla}: Of the 3,074 XCMs, we observe agreement for 1,308 XCMs and disagreement for 1,766 XCMs, with a Cohen's Kappa score of 0.22. Based on Cohen's Kappa score, the agreement level is `fair'~\citep{Landis:Koch:Kappa:Range}.}
\item{\textbf{Openstack}: Of the 7,808 XCMs, we observe agreement for 3,188 XCMs, and disagreements for 4,620 XCMs. The Cohen's Kappa score was 0.21. Based on Cohen's Kappa score, the agreement level is `fair'~\citep{Landis:Koch:Kappa:Range}.}
\item{\textbf{Wikimedia}: Of the 972 XCMs, we observe agreement for 415 XCMs, and disagreements for 557 XCMs, with a Cohen's Kappa score of 0.23. Based on Cohen's Kappa score, the agreement level is `fair'~\citep{Landis:Koch:Kappa:Range}.}
\end{itemize}

The first author of the paper was the resolver, and resolved disagreements for all four datasets. In case of disagreements the resolver's categorization is considered as final.  

We observe that the raters agreement level to be `fair' for all four datasets. One possible explanation can be that the raters agreed on whether an XCM is defect-related, but disagreed on which of the 10 defect category of the defect is related to. For defect categorization, fair or poor agreement amongst raters however, is not uncommon. Henningsson et al.~\cite{Henningsson:ISESE2004} also reported a low agreement amongst raters.

\textbf{Practitioner Agreement}: We report the agreement level between the raters' and the practitioners' categorization for randomly selected 50 XCMs as following: 
\begin{itemize} 
\item{\textbf{Mirantis}: We contact three practitioners and all of them respond. We observe a 89.0\% agreement with a Cohen's Kappa score of 0.8. Based on Cohen's Kappa score, the agreement level is `substantial'~\citep{Landis:Koch:Kappa:Range}.}
\item{\textbf{Mozilla}:   We contact six practitioners and all of them respond. We observe a 94.0\% agreement with a Cohen's Kappa score of 0.9. Based on Cohen's Kappa score, the agreement level is `almost perfect'~\citep{Landis:Koch:Kappa:Range}.}
\item{\textbf{Openstack}: We contact 10 practitioners and all of them respond. We observe a 92.0\% agreement with a Cohen's Kappa score of 0.8. Based on Cohen's Kappa score, the agreement level is `substantial'~\citep{Landis:Koch:Kappa:Range}.}
\item{\textbf{Wikimedia}: We contact seven practitioners and all of them respond. We observe a 98.0\% agreement with a Cohen's Kappa score of 0.9. Based on Cohen's Kappa score, the agreement level is `almost perfect'~\citep{Landis:Koch:Kappa:Range}.}
\end{itemize}
}
\end{itemize}

We observe that the agreement between ours and the practitioners' categorization varies from 0.8 to 0.9, which is higher than that of the agreement between the raters in the Categorization Phase. One possible explanation can be related to how the resolver resolved the disagreements. The first author of the paper has industry experience in writing IaC scripts, which may help to determine categorizations that are consistent with practitioners. Another possible explanation can be related to the sample provided to the practitioners. The provided sample, even though randomly selected, may include commit messages whose categorization are relatively easy to agree upon.



\section{Empirical Findings}
\label{results}

We report our findings in this section. 


\subsection{RQ-1: What source code properties characterize defective infrastructure as code scripts?}
\label{res-rq1}

We use the 558 defect-related commits collected from the Mozilla dataset to identify source code properties of IaC scripts that correlate with defects. By applying constructivist grounded theory described in Section~\ref{meth-rq1} we identify 12 properties of IaC scripts. Each of these properties are listed in Table~\ref{table-metrics} in the `Property' column. A brief description of each identified property is listed in the `Description' column. 


\begin{table*}[]
\captionsetup{justification=centering}
\caption{Source Code Properties that Characterize Defective IaC Scripts}
\label{table-metrics}
{\footnotesize     
\begin{tabular}{|p{1.5cm}|p{7.5cm}|p{4.5cm}|}
\hline
\textbf{Property} & \textbf{Description} & \textbf{Measurement Technique} \\
\hline
Attribute         & Attributes are code properties where configuration values are specified using the `$=>$' sign & Total count of `$=>$' usages \\ 
\hline
Command           & Commands are source code properties that are used to execute bash and batch commands  & Count of `cmd' syntax occurrences \\ 
\hline 
Comment           & Comments are non-executable parts of the script that are used for explanation & Total count of comments\\ 
\hline 
Ensure            & Ensure is a source code property that is used to check the existence of a file  & Count of `ensure' syntax occurrences\\ 
\hline 
File              & File is a source code property used to manage files, directories, and symbolic links & Count of `file' syntax occurrences\\ 
\hline
File mode         & File mode is a source code property used to set permissions of files & Count of `mode' syntax occurrences\\ 
\hline
Hard-coded string & Configuration values specified as hard-coded strings & Count of string occurrences\\
\hline
Include           & Include is a source code property that is used to execute other Puppet modules and scripts & Count of `include' syntax occurrences\\ 
\hline
Lines of code     & Size of scripts as measured by lines of code can contribute to defects & Total lines of code\\ 
\hline 
Require           & Require is a function that is used to apply resources declared in other scripts & Count of `require' syntax occurrences\\ 
\hline
SSH\_KEY          & SSH\_KEY is a source code property that sets and updates ssh keys for users & Count of `ssh\_authorized\_key' syntax occurrences\\ 
\hline
URL               & URL refers to URLs used to specify a configuration  & Count of URL occurrences\\ 
\hline
\end{tabular}
}
\end{table*}  


The median values of 12 source code properties for both defective and non-defective scripts in presented in Table~\ref{tab-res-rq1-dist}. The `D' and `ND' respectively presents the median values of each property for defective and non-defective scripts. For example, in the case of the Mirantis, the median values for attribute is respectively 23.0 and 6.5.  

 
In Table~\ref{res-table-prop-stats}, for each property we report the p-value and Cliff's Delta values respectively in the `p-value' and `Cliff' columns. We observe 10 of the 12 identified properties to show correlation with defective IaC scripts for all four datasets. The Cliff's Delta value is `large' for lines of code for three of the four datasets. The property hard-coded string has a `large' Cliff's Delta value for Mirantis and Wikimedia.   

\begin{table*}[]
\captionsetup{justification=centering}
\caption{Median values of 12 source code properties for both defective and non-defective scripts.}
\label{tab-res-rq1-dist}  
\begin{tabular}{|p{2.5cm} | p{1.2cm} p{1.2cm} | p{1.2cm} p{1.2cm} | p{1.2cm} p{1.2cm} | p{1.2cm} p{1.2cm}|}
\hline
\textbf{Property} & \multicolumn{2}{c|}{\textbf{Mirantis}} & \multicolumn{2}{c|}{\textbf{Mozilla}} & \multicolumn{2}{c|}{\textbf{Openstack}} & \multicolumn{2}{c|}{\textbf{Wikimedia}}\\
 & \textbf{D} & \textbf{ND} & \textbf{D} & \textbf{ND} & \textbf{D} & \textbf{ND} & \textbf{D} & \textbf{ND}\\
\hline
Attribute          & 23.0 & 6.5 & 10.0  & 3.0 & 13.0 & 5.0 & 12.0 & 3.0\\  
Comment            & 14.0 & 4.5 & 3.0  & 3.0 & 17.0 & 21.0 & 8.0 & 4.0\\ 
Command            & 0.0 & 0.0 & 0.0  & 0.0 & 0.0 & 0.0 & 0.0 & 0.0\\ 
Ensure             & 1.0 & 1.0 & 1.0  & 1.0 & 0.0 & 0.0 & 1.0 & 0.0\\ 
File               & 1.0 & 0.0 & 1.0  & 0.0 & 0.0 & 0.0 & 1.0 & 0.0\\  
File mode          & 1.0 & 0.0 & 0.0  & 0.0 & 0.0 & 0.0 & 0.0 & 0.0\\  
Hard-coded string  & 19.5 & 4.0 & 4.0  & 2.0 & 8.0 & 4.0 & 8.0 & 2.0\\ 
Include            & 5.0 & 1.0 & 4.0  & 2.0 & 2.0 & 1.0 & 4.0 & 1.0\\  
Lines of Code      & 90.0 & 38.0 & 53.0  & 25.0 & 77.0 & 46.0 & 57.0 & 20.0\\ 
Require            & 1.0 & 0.0 & 0.0  & 0.0 & 0.0 & 0.0 & 1.0 & 0.0\\  
SSH\_KEY           & 1.0 & 0.0 & 0.0  & 0.0 & 0.0 & 0.0 & 0.0 & 0.0\\  
URL                & 0.0 & 0.0  & 1.0  & 1.0 & 0.0 & 0.0 & 0.0 & 0.0\\ 
\hline
\end{tabular}
\end{table*}  

\begin{table*}[]
\captionsetup{justification=centering}
\caption{Validation of identified source code properties. Highlighted cells in grey indicate properties for which p-value $< 0.05$ for all four datasets.}
\label{res-table-prop-stats}  
\footnotesize
{
\begin{tabular}{| p{2.0cm} | p{1.2cm} p{0.6cm} | p{1.2cm} p{0.6cm} | p{1.2cm} p{0.6cm} | p{1.2cm} p{0.6cm} | }
\hline
\textbf{Property} & \multicolumn{2}{c|}{\textbf{Mirantis}} & \multicolumn{2}{c|}{\textbf{Mozilla}} & \multicolumn{2}{c|}{\textbf{Openstack}} & \multicolumn{2}{c|}{\textbf{Wikimedia}} \\
\hline
 & \textbf{p-value} & \textbf{Cliff} & \textbf{p-value} & \textbf{Cliff} & \textbf{p-value} & \textbf{Cliff} & \textbf{p-value} & \textbf{Cliff}\\
\hline
\cellcolor{lightgray} Attribute & $< 0.001$  & 0.47  & $< 0.001$ & 0.40 & $< 0.001$ & 0.34 & $< 0.001$ & 0.47\\ 
\hline
\cellcolor{lightgray}Command  & $< 0.001$  & 0.24 & $< 0.001$ & 0.18 & $< 0.001$ & 0.06 & $< 0.001$ & 0.18\\ 
\hline 
Comment & $< 0.001$  & 0.36 & 0.23 & 0.02 & 0.43 & 0.00 & $< 0.001$ & 0.22\\ 
\hline 
\cellcolor{lightgray}Ensure & $< 0.001$  & 0.38 & 0.02 & 0.09 & $< 0.001$ & 0.19 & $< 0.001$ & 0.28\\ 
\hline 
\cellcolor{lightgray}File &  $< 0.001$  & 0.36 & $< 0.001$ & 0.18 & $< 0.001$ & 0.08 & $< 0.001$ & 0.31\\ 
\hline
\cellcolor{lightgray}File mode & $< 0.001$  & 0.40  & $< 0.001$ & 0.24 & $< 0.001$ & 0.06 & $< 0.001$ & 0.23\\ 
\hline
\cellcolor{lightgray} Hard-coded string & $< 0.001$  & 0.55  & $< 0.001$ & 0.40 & $< 0.001$ & 0.37 & $< 0.001$ & 0.54\\
\hline
\cellcolor{lightgray}Include & $< 0.001$  & 0.32  & $< 0.001$ & 0.31 & $< 0.001$ & 0.22 & $< 0.001$ & 0.37\\ 
\hline
\cellcolor{lightgray}Lines of code & $< 0.001$  & 0.50 & $< 0.001$  & 0.51 & $< 0.001$ & 0.32 & $< 0.001$ & 0.51\\ 
\hline 
\cellcolor{lightgray}Require & $< 0.001$  & 0.35  & $< 0.001$ & 0.19 & $< 0.001$ & 0.11 & $< 0.001$ & 0.32\\ 
\hline
\cellcolor{lightgray} SSH\_KEY & $< 0.001$  & 0.39  & $< 0.001$ & 0.24 & $< 0.001$ & 0.07 & $< 0.001$ & 0.24\\ 
\hline
URL  & $< 0.001$  & 0.22 & 0.009  & 0.08 & 0.40 & 0.00 & $< 0.001$ & 0.17\\ 
\hline
\end{tabular}
}
\end{table*}  


We report the feature importance values for each identified source code property in Table~\ref{res-table-rf}. For three datasets, we observe lines of code to show strongest correlation with defective scripts. For Mirantis, we observe the strongest correlation to be hard-coded string. 

\begin{sidewaystable}[]
\captionsetup{justification=centering}
\caption{Ranked order of the 12 source code properties that show highest correlation according to feature importance analysis. The `Practitioner Agreement' column presents the percentage of practitioners who agreed or strongly agreed with the property.}
\label{res-table-rf}
\footnotesize
{
\begin{tabular}{p{0.75cm} p{2.9cm} p{2.9cm} p{2.9cm} p{2.9cm} | p{3.5cm}}
\hline
\textbf{Rank}  & \textbf{Mirantis} & \textbf{Mozilla} & \textbf{Openstack} & \textbf{Wikimedia}  & \textbf{Properties Practitioners Agreed With} \\
\hline
1  & Hard-coded string (0.20)  & Lines of code (0.27)      & Lines of code (0.26)     & Lines of code (0.19)     & Include (62\%)\\ 
2  & Lines of code (0.17)      & Attribute (0.17)          & Hard-coded string (0.18) & Attribute (0.17)         & Hard-coded string (58\%)\\ 
3  & Command (0.11)            & Hard-coded string (0.15)  & Attribute (0.15)         & Hard-coded string (0.13) & URL (52\%) \\ 
4  & Comment (0.11)            & Include (0.14)            & Comment (0.14)           & Comment (0.11)           & Command (46\%)\\ 
5  & Attribute (0.10)          & Comment (0.05)            & Include (0.09)           & Include (0.10)           & Lines of code (42\%)\\ 
6  & File mode (0.08)          & Ensure (0.04)             & URL (0.04)               & File (0.08)              & Require (42\%)\\ 
7  & Require (0.07)            & File (0.03)               & Ensure (0.03)            & Ensure (0.05)            & File (31\%) \\
8  & Ensure (0.06)             & Require (0.03)            & Command (0.02)           & Require (0.05)           & Attribute (27\%) \\ 
9  & Include (0.06)            & File mode (0.02)          & File (0.02)              & URL (0.03)               & Comment (23\%) \\ 
10 & URL (0.03)                & Command (0.02)            & Require (0.02)           & Command (0.02)           & SSH\_KEY (23\%) \\ 
11 & SSH\_KEY (0.02)           & URL (0.02)                & File mode (0.01)         & File mode (0.01)         & Ensure (19\%) \\ 
12 & File (0.01)               & SSH\_KEY (0.01)           & SSH\_KEY (0.01)          & SSH\_KEY (0.00)          & File mode (15\%)\\ 
\hline
\end{tabular}
}
\end{sidewaystable} 

Our findings related to feature importance is in congruence with our findings presented in Table~\ref{res-table-prop-stats}. Cliff's delta value is `large' for lines of code for three datasets. Our feature importance analysis identifies lines of code as the property with the strongest correlation for three datasets. According to Table~\ref{res-table-rf}, hard-coded string is identified as the strongest correlating property and also has a `large' Cliff's Delta value for Mirantis.    


\subsection{RQ-2: Do practitioners agree with the identified source code properties?}
\label{res-rq2}

As mentioned in Section~\ref{meth-rq2} we conduct a survey with 350 practitioners to quantify if practitioners agreed with our set of 12 source code properties. We obtain a survey response rate of 7.4\% (26 out of 350). The reported experience level in Puppet is listed in Table~\ref{res-rq2-tab-exp}. The `Experience' column lists the categories for experience in Puppet. The `Count' column presents the number of practitioners who identified with the corresponding experience level.  

Of the 12 properties, practitioners showed the highest agreement with include, in contrary to our feature importance analysis. The least agreed property is `File mode'. Reported agreement level by all practitioners is presented in Table~\ref{fig-res-rq2}. For three properties we observe at least 50\% of the practitioners to agree with. These three properties are: URL, Hard-coded string, and Include.    

We also compare practitioner survey responses and our feature importance analysis by presenting the practitioner agreement level in Table~\ref{res-table-rf}. We also report the percentage of practitioners who agreed or strongly agreed with a certain property in the `Practitioner Agreement' column. According to survey results, majority of the practitioners agreed with include in contrary to the feature importance analysis for the four datasets.    

\begin{table*}[]
\captionsetup{justification=centering}
\caption{Survey responses from practitioners. Of the 12 properties majority of the practitioners agreed with include.} 
\label{fig-res-rq2}  
\begin{tabular}{p{3.2cm}  p{2.3cm} p{2.0cm}  p{2.0cm} p{1.8cm} p{2.0cm} }
\hline
 & \textbf{Strongly disagree (\%)} & \textbf{Disagree (\%)} & \textbf{Neutral (\%)} & \textbf{Agree (\%)} & \textbf{Strongly agree (\%)} \\
\hline
Attribute          & 11.5 & 26.9 & 34.6  & 23.0 & 3.8 \\  
Comment            & 15.3 & 38.4 & 23.0  & 15.3 & 7.7 \\ 
Command            & 3.8  & 3.8  & 46.1  & 30.7 & 15.3 \\ 
Ensure             & 3.8  & 38.4 & 38.4  & 7.7  & 11.53 \\ 
File               & 3.8  & 15.3 & 50.0  & 23.0 & 7.7 \\  
File mode          & 3.8  & 19.2 & 61.5  & 11.5 & 3.8 \\  
Hard-coded string  & 3.8  & 11.5 & 26.9  & 46.1 & 11.5 \\ 
Include            & 3.8  & 11.5 & 23.0  & 46.1 & 15.3 \\  
Lines of Code      & 7.7  & 15.3 & 34.6  & 34.6 & 7.7 \\ 
Require            & 3.8  & 23.0 & 30.7  & 34.6 & 7.7 \\  
SSH\_KEY           & 3.8  & 15.3 & 57.7  & 15.3 & 7.7 \\  
URL                & 3.8  & 11.5 & 30.7  & 46.1 & 3.8 \\ 
\hline
\end{tabular}
\end{table*} 


\begin{table}[]
\centering
\caption{Reported practitioner experience in Puppet script development}
\label{res-rq2-tab-exp}
\begin{tabular}{ p{5.0cm}  p{5.0cm} }
\hline
\textbf{Experience (Years)}   & \textbf{Count} \\
\hline
\textbf{$<$ 1 } &  1 (3.9\%)\\
\textbf{1-2} &  6 (23.0\%) \\
\textbf{3-5} & 11 (42.3\%) \\
\textbf{6-10} & 7 (26.9\%) \\
\textbf{$>$ 10} & 1 (3.9\%) \\
\hline
\end{tabular}
\end{table}

On the contrary to our statistical analysis, we observe practitioners to show highest agreement with include. One possible explanation can be attributed to practitioner perception. Devanbu et al.~\cite{Devanbu:Belief:Evidence} reported that practitioner's perceptions can be incongruent with empirical data. Our findings provide further evidence that empirical results can be incongruent with practitioner's perceptions.  


Despite the disagreements between our empirical findings and practitioner responses, our findings can be helpful. Our findings can inform practitioners on the existence of source code properties that require sufficient inspection and testing efforts. Based on our findings practitioners can benefit from rigorous inspection and testing when any of the 10 identified properties appear in an IaC script. 

We also observe some level of congruence between our statistical analysis and survey responses. The second most agreed upon property is hard-coded string, which is identified as the most correlated property for Mirantis. So, both based on survey data and feature importance analysis, we can conclude presence of hard-coded string in IaC scripts make scripts defect-prone.


\subsection{RQ-3: How can we construct defect prediction models for infrastructure as code scripts using the identified source code properties?}
\label{res-rq3}

As described in Section~\ref{meth-rq3}, we use PCA analysis to construct prediction models needed for RQ-3. We report the number of principal components that account for at least 95\% of the data variability in Table~\ref{res-rq3-pca}. The column `Property-based' provides the number of principal components that account for 95\% of the total variance where we used 12 source code properties to construct prediction models. For example, the number of principal components that account for at least 95\% of the data variability for the `Property-based' approach and the `Bag-of-words' approach is respectively, 1 and 50. 

The median AUC values are presented in Table~\ref{res:rq3:table:auc}. The column `Property-based' provides the median AUC values using the 12 identified properties. For AUC the property-based prediction model outperforms the bag-of-words technique for three datasets, but is tied with the bag-of-words approach for one dataset. LR provided the highest median AUC for two datasets using our 12 properties.   

We report the median precision, recall, and F-measure values for 10 $\times$ 10 cross validation, for all learners and all datasets respectively in Tables~\ref{res:rq3:table:pre},~\ref{res:rq3:table:rec}, and~\ref{res:rq3:table:f1}. The column `Property-based' provides the median AUC values using the 12 identified properties, whereas, the `Bag-of-words' column presents the median prediction performance values for the bag-of-words technique. As shown in Figure~\ref{res:rq3:table:pre}, for NB we observe the highest median precision for all four datasets, where the median precision is greater than 0.80. According to Table~\ref{res:rq3:table:rec}, CART provides the highest median recall for two datasets, whereas the highest median recall is obtained for KNN and LR respectively for Mirantis and Openstack. CART, KNN, and LR provides the highest median F-measure for two datasets according to Table~\ref{res:rq3:table:f1}. For three measures precision, recall, and F-measure, our property-based prediction models outperform the bag-of-words technique.   


\begin{table}[]
\centering
\caption{Number of Principle components used for prediction models}
\label{res-rq3-pca}
\footnotesize
{
\begin{tabular}{ p{5.0cm}  p{2.5cm} p{2.5cm} }
\hline
\textbf{Dataset}   & \textbf{Property-based} & \textbf{Bag-of-words} \\
\hline
\textbf{Mirantis}  & 1 &  50 \\
\textbf{Mozilla}   & 1 &  140\\
\textbf{Openstack} & 2 &  400\\
\textbf{Wikimedia} & 2 &  150\\
\hline
\end{tabular}
}
\end{table}

\begin{table*}[]
\centering
\caption{AUC for each model building technique. The highlighted cell in grey indicates the best technique, as determined by the Scott Knot Test.}
\label{res:rq3:table:auc}
\footnotesize{  
\begin{tabular}{| p{1.2cm} | p{0.9cm}p{0.9cm}p{0.6cm}p{0.6cm}p{0.6cm}|p{0.9cm}p{0.9cm}p{0.6cm}p{0.6cm}p{0.6cm} |}
\hline
\textbf{Dataset} & \multicolumn{5}{c|}{\textbf{Property-based}} & \multicolumn{5}{c|}{\textbf{Bag-of-words}} \\
\hline
 &  \textbf{CART} & \textbf{KNN} & \textbf{LR} & \textbf{NB} & \textbf{RF} & \textbf{CART} & \textbf{KNN} & \textbf{LR} & \textbf{NB} & \textbf{RF} \\
\hline
MIR & 0.65 & 0.67 & \cellcolor{lightgray}0.71 & 0.62 & 0.65 & 0.61 & 0.65 & 0.57 & 0.64 & 0.66  \\
MOZ & \cellcolor{lightgray} 0.71 & 0.66 & 0.69 & 0.66 & 0.69 & 0.52 & 0.48 & 0.51 & 0.60 & 0.56  \\
OST & 0.52 & 0.54 & 0.63 & \cellcolor{lightgray} 0.66 & 0.54 & 0.55 & 0.55 & 0.64 & 0.63 & 0.56  \\
WIK & 0.64 & 0.65 & \cellcolor{lightgray} 0.68 & 0.64 & 0.64 & 0.57 & 0.52 & 0.47 & \cellcolor{lightgray} 0.68 & 0.61  \\
\hline
\end{tabular}
}
\end{table*}  


\begin{table*}[]
\centering
\caption{Precision for each model building technique. The highlighted cell in grey indicates the best technique, as determined by the Scott Knot Test.}
\label{res:rq3:table:pre}
\footnotesize{  
\begin{tabular}{|p{1.2cm}|p{0.9cm}p{0.9cm}p{0.6cm}p{0.6cm}p{0.6cm}|p{0.9cm}p{0.9cm}p{0.6cm}p{0.6cm}p{0.6cm}|}
\hline
\textbf{Dataset} & \multicolumn{5}{c|}{\textbf{Property-based}} & \multicolumn{5}{c|}{\textbf{Bag-of-words}} \\
\hline
 &  \textbf{CART} & \textbf{KNN} & \textbf{LR} & \textbf{NB} & \textbf{RF} & \textbf{CART} & \textbf{KNN} & \textbf{LR} & \textbf{NB} & \textbf{RF} \\
\hline
MIR & 0.65 & 0.69 & 0.78 & \cellcolor{lightgray} 0.80 & 0.68 & 0.62 & 0.74 & 0.63 & 0.75 & 0.69  \\
MOZ & 0.68 & 0.63 & 0.73 & \cellcolor{lightgray} 0.85 & 0.67 & 0.51 & 0.41 & 0.48 & 0.39 & 0.58  \\
OST & 0.60 & 0.62 & 0.70 & \cellcolor{lightgray} 0.84 & 0.62 & 0.63 & 0.64 & 0.65 & 0.76 & 0.64  \\
WIK & 0.67 & 0.68 & 0.74 & \cellcolor{lightgray} 0.85 & 0.68 & 0.60 & 0.60 & 0.51 & 0.76 & 0.64  \\
\hline
\end{tabular}
}
\end{table*}  


\begin{table*}[]
\centering
\caption{Recall for each model building technique. The highlighted cell in grey indicates the best technique, as determined by the Scott Knot Test.}
\label{res:rq3:table:rec}
\footnotesize{  
\begin{tabular}{|p{1.2cm}|p{0.9cm}p{0.9cm}p{0.6cm}p{0.6cm}p{0.6cm}|p{0.9cm}p{0.9cm}p{0.6cm}p{0.6cm}p{0.6cm}|}
\hline
\textbf{Dataset} & \multicolumn{5}{c|}{\textbf{Property-based}} & \multicolumn{5}{c|}{\textbf{Bag-of-words}} \\
\hline
 &  \textbf{CART} & \textbf{KNN} & \textbf{LR} & \textbf{NB} & \textbf{RF} & \textbf{CART} & \textbf{KNN} & \textbf{LR} & \textbf{NB} & \textbf{RF} \\
\hline
MIR & 0.66 & \cellcolor{lightgray} 0.70 & 0.63 & 0.34 & 0.66 & 0.69 & 0.49 & 0.48 & 0.45 & 0.64  \\
MOZ & \cellcolor{lightgray} 0.66 & 0.61 & 0.54 & 0.37 & 0.64 & 0.25 & 0.22 & 0.21 & 0.39 & 0.27  \\
OST & 0.60 & 0.60 & \cellcolor{lightgray} 0.67 & 0.42 & 0.58 & 0.62 & 0.50 & 0.57 & 0.46 & 0.57  \\
WIK & \cellcolor{lightgray} 0.67 & \cellcolor{lightgray} 0.67 & 0.63 & 0.35 & 0.63 & 0.65 & 0.24 & 0.30 & 0.59 & 0.64  \\
\hline
\end{tabular}
}
\end{table*}  


\begin{table*}[]
\centering
\caption{F-measure for each model building technique. The highlighted cell in grey indicates the best technique, as determined by the Scott Knot Test.}
\label{res:rq3:table:f1}
\footnotesize{  
\begin{tabular}{|p{1.2cm}|p{0.9cm}p{0.9cm}p{0.6cm}p{0.6cm}p{0.6cm}|p{0.9cm}p{0.9cm}p{0.6cm}p{0.6cm}p{0.6cm}|}
\hline
\textbf{Dataset} & \multicolumn{5}{c|}{\textbf{Property-based}} & \multicolumn{5}{c|}{\textbf{Bag-of-words}} \\
\hline
 &  \textbf{CART} & \textbf{KNN} & \textbf{LR} & \textbf{NB} & \textbf{RF} & \textbf{CART} & \textbf{KNN} & \textbf{LR} & \textbf{NB} & \textbf{RF} \\
\hline
MIR & 0.67 & \cellcolor{lightgray} 0.70 & \cellcolor{lightgray} 0.70 & 0.48 & 0.67 & 0.66 & 0.59 & 0.64 & 0.63 & 0.67  \\
MOZ & \cellcolor{lightgray} 0.67 & 0.62 & 0.62 & 0.52 & 0.65 & 0.34 & 0.29 & 0.29 & 0.48 & 0.37  \\
OST & 0.60 & 0.61 & \cellcolor{lightgray} 0.68 & 0.56 & 0.60 & 0.62 & 0.56 & 0.61 & 0.58 & 0.60  \\
WIK & \cellcolor{lightgray} 0.67 & \cellcolor{lightgray} 0.67 & 0.68 & 0.50 & 0.66 & 0.63 & 0.35 & 0.38 & 0.66 & 0.65  \\
\hline
\end{tabular}
}
\end{table*}  



\section{Discussion}
\label{discussion}
We discuss our findings with possible implications as following:
\subsection{Implications for Practitioners}

\textbf{Prioritization of Inspection Efforts}: Our findings have implications on how practitioners can prioritize inspection efforts for IaC scripts. The identified 12 source code properties can be helpful in early prediction of defective scripts. As shown in Tables~\ref{res-table-prop-stats} and~\ref{res-table-rf}, hard-coded string is correlated with making defective IaC scripts, and therefore, test cases can be designed by focusing on string-related values assigned in IaC scripts. 

Code inspection efforts can also be prioritized using our findings. According to our feature importance analysis, attribute is correlated with defective IaC scripts. IaC scripts with relatively large amount of attributes can get extra scrutiny. From Table~\ref{res-table-rf} we observe other IaC-related source code properties that contribute to defective IaC scripts. Examples of such properties include: setting a file path (File), and executing external modules or scripts (Include). Practitioners might benefit from code inspection using manual peer reviews for these particular properties, as well.   

\textbf{Tools}: Prior research~\cite{ICSE2013:BugPredictionForHuman} observed that defect prediction models can be helpful for programmers who write code in general prupose programming languages. Defect prediction of software artifacts is now offered as a cloud-service, as done by DevOps Insights~\footnote{https://www.ibm.com/cloud/devops-insights}. For IaC scripts we observe the opportunity of creating a new set of tools and services that will help in defect mitigation. Toolsmiths can use our prediction models to build tools that pinpoint the defective IaC scripts that need to be fixed. Such tools can explicitly state which source code properties are more correlated with defects than others and need special attention when making changes. 

\subsection{Future Research}
Our paper provides opportunity for further research in the area of defect prediction of IaC scripts. Sophisticated statistical techniques, such as topic modeling and deep learning, can be applied to discover more IaC-related source code properties. We have not accounted for development activity metrics such as number of programmers who modified an IaC script. Future research can investigate the applicability of development activity metrics. Researchers can also investigate how practitioners in real life perceive and use defect prediction models for IaC scripts.


\section{Threats to Validity}
\label{threats}

We discuss the limitations of our paper as following:

\begin{itemize}

\item{\textbf{Conclusion Validity}: Our approach is based on qualitative analysis, where raters categorized XCMs, and assigned defect categories. We acknowledge that the process is susceptible human judgment, and the raters' experience can bias the categories assigned. The accompanying human subjectivity can influence the distribution of the defect category for IaC scripts of interest. We mitigated this threat by assigning multiple raters for the same set of XCMs. Next, we used a resolver, who resolved the disagreements. Further, we cross-checked our categorization with practitioners who authored the XCMs, and observed `substantial' to `almost perfect' agreement. 

For RQ-2 the survey response rate was 7.4\%. We acknowledge that the survey response rate was low, and our findings may not be generalizable. 
}

\item{\textbf{Internal Validity}: We have used a combination of commit messages and issue report descriptions to determine if an IaC script is associated with a defect. We acknowledge that these messages might not have given the full context for the raters. Other sources of information such as practitioner input, and code changes that take place in each commit could have provided the raters better context to categorize the XCMs. 

We also acknowledge that our set of properties is not comprehensive. We derived these properties by applying qualitative analysis on defect-related commits of one dataset. We mitigated this limitation by applying empirical analysis on three more datasets, and quantify if the identified properties show correlation with defective scripts.   

}

\item{\textbf{Construct validity}: Our process of using human raters to determine defect categories can be limiting, as the process is susceptible to mono-method bias, where subjective judgment of raters can influence the findings. We mitigated this threat by using multiple raters.  

Also, for Mirantis and Wikimedia, we used graduate students who performed the categorization as part of their class work. Students who participated in the categorization process can be subject to evaluation apprehension, i.e. consciously or sub-consciously relating their performance with the grades they would achieve for the course. We mitigated this threat by clearly explaining to the students that their performance in the categorization process would not affect their grades. 

The raters involved in the categorization process had professional experience in software engineering for at two years on average. Their experience in software engineering may make the raters curious about the expected outcomes of the categorization process, which may effect the distribution of the categorization process. Furthermore, the resolver also has professional experience in software engineering and IaC script development, which could influence the outcome of the defect category distribution. 
}

\item{\textbf{External Validity}: Our scripts are collected from the OSS domain, and not from proprietary sources. Our findings are subject to external validity, as our findings may not be generalizable.

We construct our datasets using Puppet, which is a declarative language. Our findings may not generalize for IaC scripts that use an imperative form of language.  
} 

\end{itemize}


\section{Conclusion}
\label{conclusion}

In continuous deployment, IT organizations rapidly deploy software and services to end-users using an automated deployment pipeline. IaC is a fundamental pillar to implement an automated deployment pipeline. Defective IaC scripts can hinder the reliability of the automated deployment pipeline. Characterizing source code properties of IaC scripts that correlate with defective IaC scripts can help identify signals to increase the quality of IaC scripts. We apply qualitative analysis to identify 12 source code properties that correlate with defective IaC scripts. We observe 10 of the 12 properties to show correlation with defective IaC scripts for all four datasets. The properties that show the strongest correlation are lines of code and hard-coded string. In contrast to our empirical analysis, we observe practitioners to agree most with the `URL' property. Using our 12 properties we construct defect prediction models, which outperform the bag-of-words technique with respect to precision, recall, and F-measure. We hope our paper will facilitate further research in the area of defect analysis for IaC scripts.

\section*{Acknowledgments}
We thank the practitioners for responding to our e-mails related to defect categorization and survey analysis. We also thank the members of the RealSearch group for their valuable feedback. 

\section{References}
\bibliographystyle{elsarticle-num} 
\bibliography{ist}

\begin{thebibliography}{10}
\expandafter\ifx\csname url\endcsname\relax
  \def\url#1{\texttt{#1}}\fi
\expandafter\ifx\csname urlprefix\endcsname\relax\def\urlprefix{URL }\fi
\expandafter\ifx\csname href\endcsname\relax
  \def\href#1#2{#2} \def\path#1{#1}\fi

\bibitem{me:agile:cd2015}
A.~A.~U. Rahman, E.~Helms, L.~Williams, C.~Parnin,
  \href{http://dx.doi.org/10.1109/Agile.2015.12}{Synthesizing continuous
  deployment practices used in software development}, in: Proceedings of the
  2015 Agile Conference, AGILE '15, IEEE Computer Society, Washington, DC, USA,
  2015, pp. 1--10.
\newblock \href {https://doi.org/10.1109/Agile.2015.12}
  {\path{doi:10.1109/Agile.2015.12}}.
\newline\urlprefix\url{http://dx.doi.org/10.1109/Agile.2015.12}

\bibitem{Humble:2010:CD}
J.~Humble, D.~Farley, Continuous Delivery: Reliable Software Releases Through
  Build, Test, and Deployment Automation, 1st Edition, Addison-Wesley
  Professional, 2010.

\bibitem{parnin:adages}
C.~Parnin, E.~Helms, C.~Atlee, H.~Boughton, M.~Ghattas, A.~Glover, J.~Holman,
  J.~Micco, B.~Murphy, T.~Savor, M.~Stumm, S.~Whitaker, L.~Williams, The top 10
  adages in continuous deployment, IEEE Software 34~(3) (2017) 86--95.
\newblock \href {https://doi.org/10.1109/MS.2017.86}
  {\path{doi:10.1109/MS.2017.86}}.

\bibitem{me:icst2018:iac}
A.~Rahman, L.~Williams, Characterizing defective configuration scripts used for
  continuous deployment, in: 2018 IEEE 11th International Conference on
  Software Testing, Verification and Validation (ICST), 2018, pp. 34--45.
\newblock \href {https://doi.org/10.1109/ICST.2018.00014}
  {\path{doi:10.1109/ICST.2018.00014}}.

\bibitem{Cito:Cloud2015:FSE}
J.~Cito, P.~Leitner, T.~Fritz, H.~C. Gall,
  \href{http://doi.acm.org/10.1145/2786805.2786826}{The making of cloud
  applications: An empirical study on software development for the cloud}, in:
  Proceedings of the 2015 10th Joint Meeting on Foundations of Software
  Engineering, ESEC/FSE 2015, ACM, New York, NY, USA, 2015, pp. 393--403.
\newblock \href {https://doi.org/10.1145/2786805.2786826}
  {\path{doi:10.1145/2786805.2786826}}.
\newline\urlprefix\url{http://doi.acm.org/10.1145/2786805.2786826}

\bibitem{JiangAdamsMSR2015}
Y.~Jiang, B.~Adams,
  \href{http://dl.acm.org/citation.cfm?id=2820518.2820527}{Co-evolution of
  infrastructure and source code: An empirical study}, in: Proceedings of the
  12th Working Conference on Mining Software Repositories, MSR '15, IEEE Press,
  Piscataway, NJ, USA, 2015, pp. 45--55.
\newline\urlprefix\url{http://dl.acm.org/citation.cfm?id=2820518.2820527}

\bibitem{cart:original}
L.~Breiman, et~al.,
  \href{http://www.crcpress.com/catalog/C4841.htm}{{Classification and
  Regression Trees}}, 1st Edition, Chapman \& Hall, New York, 1984.
\newline\urlprefix\url{http://www.crcpress.com/catalog/C4841.htm}

\bibitem{Tan:2005:IDM}
P.-N. Tan, M.~Steinbach, V.~Kumar, Introduction to Data Mining, (First
  Edition), Addison-Wesley Longman Publishing Co., Inc., Boston, MA, USA, 2005.

\bibitem{logit:original}
D.~Freedman, Statistical Models : Theory and Practice, {Cambridge University
  Press}, 2005.

\bibitem{Breiman2001:RF:ORIGINAL}
L.~Breiman, \href{http://dx.doi.org/10.1023/A:1010933404324}{Random forests},
  Machine Learning 45~(1) (2001) 5--32.
\newblock \href {https://doi.org/10.1023/A:1010933404324}
  {\path{doi:10.1023/A:1010933404324}}.
\newline\urlprefix\url{http://dx.doi.org/10.1023/A:1010933404324}

\bibitem{harris:bag:of:words:orig}
Z.~S. Harris, Distributional structure, WORD 10~(2-3) (1954) 146--162.
\newblock \href {https://doi.org/10.1080\/00437956.1954.11659520}
  {\path{doi:10.1080\/00437956.1954.11659520}}.

\bibitem{ShambaughRehearsal2016}
R.~Shambaugh, A.~Weiss, A.~Guha,
  \href{http://doi.acm.org/10.1145/2980983.2908083}{Rehearsal: A configuration
  verification tool for puppet}, SIGPLAN Not. 51~(6) (2016) 416--430.
\newblock \href {https://doi.org/10.1145/2980983.2908083}
  {\path{doi:10.1145/2980983.2908083}}.
\newline\urlprefix\url{http://doi.acm.org/10.1145/2980983.2908083}

\bibitem{propuppet:book}
J.~T. McCune, Jeffrey,
  \href{https://www.springer.com/gp/book/9781430230571}{Pro Puppet}, 1st
  Edition, Apress, 2011.
\newblock \href {https://doi.org/10.1007/978-1-4302-3058-8}
  {\path{doi:10.1007/978-1-4302-3058-8}}.
\newline\urlprefix\url{https://www.springer.com/gp/book/9781430230571}

\bibitem{puppet-doc}
P.~Labs, {Puppet Documentation}, \url{https://docs.puppet.com/}, [Online;
  accessed 10-October-2017] (2017).

\bibitem{SharmaPuppet2016}
T.~Sharma, M.~Fragkoulis, D.~Spinellis,
  \href{http://doi.acm.org/10.1145/2901739.2901761}{Does your configuration
  code smell?}, in: Proceedings of the 13th International Conference on Mining
  Software Repositories, MSR '16, ACM, New York, NY, USA, 2016, pp. 189--200.
\newblock \href {https://doi.org/10.1145/2901739.2901761}
  {\path{doi:10.1145/2901739.2901761}}.
\newline\urlprefix\url{http://doi.acm.org/10.1145/2901739.2901761}

\bibitem{Hanappi:2016:pupp:converge}
O.~Hanappi, W.~Hummer, S.~Dustdar,
  \href{http://doi.acm.org/10.1145/3022671.2984000}{Asserting reliable
  convergence for configuration management scripts}, SIGPLAN Not. 51~(10)
  (2016) 328--343.
\newblock \href {https://doi.org/10.1145/3022671.2984000}
  {\path{doi:10.1145/3022671.2984000}}.
\newline\urlprefix\url{http://doi.acm.org/10.1145/3022671.2984000}

\bibitem{Weiss:Tortoise}
A.~Weiss, A.~Guha, Y.~Brun,
  \href{http://dl.acm.org/citation.cfm?id=3155562.3155641}{Tortoise:
  Interactive system configuration repair}, in: Proceedings of the 32Nd
  IEEE/ACM International Conference on Automated Software Engineering, ASE
  2017, IEEE Press, Piscataway, NJ, USA, 2017, pp. 625--636.
\newline\urlprefix\url{http://dl.acm.org/citation.cfm?id=3155562.3155641}

\bibitem{Hummer:IaC}
W.~Hummer, F.~Rosenberg, F.~Oliveira, T.~Eilam,
  \href{http://doi.acm.org/10.1145/2541614.2541632}{Automated testing of chef
  automation scripts}, in: Proceedings Demo:38; Poster Track of ACM/IFIP/USENIX
  International Middleware Conference, MiddlewareDPT '13, ACM, New York, NY,
  USA, 2013, pp. 4:1--4:2.
\newblock \href {https://doi.org/10.1145/2541614.2541632}
  {\path{doi:10.1145/2541614.2541632}}.
\newline\urlprefix\url{http://doi.acm.org/10.1145/2541614.2541632}

\bibitem{Bent:Saner2018:Puppet}
E.~van~der Bent, J.~Hage, J.~Visser, G.~Gousios, How good is your puppet? an
  empirically defined and validated quality model for puppet, in: 2018 IEEE
  25th International Conference on Software Analysis, Evolution and
  Reengineering (SANER), 2018, pp. 164--174.
\newblock \href {https://doi.org/10.1109/SANER.2018.8330206}
  {\path{doi:10.1109/SANER.2018.8330206}}.

\bibitem{Rahman:RCOSE17}
A.~Rahman, A.~Partho, D.~Meder, L.~Williams,
  \href{https://doi.org/10.1109/RCoSE.2017..8}{Which factors influence
  practitioners' usage of build automation tools?}, in: Proceedings of the 3rd
  International Workshop on Rapid Continuous Software Engineering, RCoSE '17,
  IEEE Press, Piscataway, NJ, USA, 2017, pp. 20--26.
\newblock \href {https://doi.org/10.1109/RCoSE.2017..8}
  {\path{doi:10.1109/RCoSE.2017..8}}.
\newline\urlprefix\url{https://doi.org/10.1109/RCoSE.2017..8}

\bibitem{Rahman:RCOSE18}
A.~Rahman, A.~Partho, P.~Morrison, L.~Williams,
  \href{http://doi.acm.org/10.1145/3194760.3194769}{What questions do
  programmers ask about configuration as code?}, in: Proceedings of the 4th
  International Workshop on Rapid Continuous Software Engineering, RCoSE '18,
  ACM, New York, NY, USA, 2018, pp. 16--22.
\newblock \href {https://doi.org/10.1145/3194760.3194769}
  {\path{doi:10.1145/3194760.3194769}}.
\newline\urlprefix\url{http://doi.acm.org/10.1145/3194760.3194769}

\bibitem{Nagappan:ICSE2005}
N.~Nagappan, T.~Ball, \href{http://doi.acm.org/10.1145/1062455.1062514}{Use of
  relative code churn measures to predict system defect density}, in:
  Proceedings of the 27th International Conference on Software Engineering,
  ICSE '05, ACM, New York, NY, USA, 2005, pp. 284--292.
\newblock \href {https://doi.org/10.1145/1062455.1062514}
  {\path{doi:10.1145/1062455.1062514}}.
\newline\urlprefix\url{http://doi.acm.org/10.1145/1062455.1062514}

\bibitem{Zheng:Laurie2006}
J.~Zheng, L.~Williams, N.~Nagappan, W.~Snipes, J.~P. Hudepohl, M.~A. Vouk,
  \href{http://dx.doi.org/10.1109/TSE.2006.38}{On the value of static analysis
  for fault detection in software}, IEEE Trans. Softw. Eng. 32~(4) (2006)
  240--253.
\newblock \href {https://doi.org/10.1109/TSE.2006.38}
  {\path{doi:10.1109/TSE.2006.38}}.
\newline\urlprefix\url{http://dx.doi.org/10.1109/TSE.2006.38}

\bibitem{Zimmermann:Eclipse2007}
T.~Zimmermann, R.~Premraj, A.~Zeller,
  \href{http://dx.doi.org/10.1109/PROMISE.2007.10}{Predicting defects for
  eclipse}, in: Proceedings of the Third International Workshop on Predictor
  Models in Software Engineering, PROMISE '07, IEEE Computer Society,
  Washington, DC, USA, 2007, pp. 9--.
\newblock \href {https://doi.org/10.1109/PROMISE.2007.10}
  {\path{doi:10.1109/PROMISE.2007.10}}.
\newline\urlprefix\url{http://dx.doi.org/10.1109/PROMISE.2007.10}

\bibitem{ieee:def}
IEEE, Ieee standard classification for software anomalies, IEEE Std 1044-2009
  (Revision of IEEE Std 1044-1993) (2010) 1--23\href
  {https://doi.org/10.1109/IEEESTD.2010.5399061}
  {\path{doi:10.1109/IEEESTD.2010.5399061}}.

\bibitem{MunaiahCuration2017}
N.~Munaiah, S.~Kroh, C.~Cabrey, M.~Nagappan,
  \href{http://dx.doi.org/10.1007/s10664-017-9512-6}{Curating github for
  engineered software projects}, Empirical Software Engineering (2017)
  1--35\href {https://doi.org/10.1007/s10664-017-9512-6}
  {\path{doi:10.1007/s10664-017-9512-6}}.
\newline\urlprefix\url{http://dx.doi.org/10.1007/s10664-017-9512-6}

\bibitem{Ray:NaturalnessICSE2016}
B.~Ray, V.~Hellendoorn, S.~Godhane, Z.~Tu, A.~Bacchelli, P.~Devanbu,
  \href{http://doi.acm.org/10.1145/2884781.2884848}{On the "naturalness" of
  buggy code}, in: Proceedings of the 38th International Conference on Software
  Engineering, ICSE '16, ACM, New York, NY, USA, 2016, pp. 428--439.
\newblock \href {https://doi.org/10.1145/2884781.2884848}
  {\path{doi:10.1145/2884781.2884848}}.
\newline\urlprefix\url{http://doi.acm.org/10.1145/2884781.2884848}

\bibitem{Zhang:Mockus2016}
F.~Zhang, A.~Mockus, I.~Keivanloo, Y.~Zou,
  \href{http://dx.doi.org/10.1007/s10664-015-9396-2}{Towards building a
  universal defect prediction model with rank transformed predictors},
  Empirical Softw. Engg. 21~(5) (2016) 2107--2145.
\newblock \href {https://doi.org/10.1007/s10664-015-9396-2}
  {\path{doi:10.1007/s10664-015-9396-2}}.
\newline\urlprefix\url{http://dx.doi.org/10.1007/s10664-015-9396-2}

\bibitem{Hassan:TSE2016:AggregateMetrics}
F.~Zhang, A.~E. Hassan, S.~McIntosh, Y.~Zou, The use of summation to aggregate
  software metrics hinders the performance of defect prediction models, IEEE
  Transactions on Software Engineering 43~(5) (2017) 476--491.
\newblock \href {https://doi.org/10.1109/TSE.2016.2599161}
  {\path{doi:10.1109/TSE.2016.2599161}}.

\bibitem{maletic:commit:icpc2008}
A.~Alali, H.~Kagdi, J.~I. Maletic, What's a typical commit? a characterization
  of open source software repositories, in: 2008 16th IEEE International
  Conference on Program Comprehension, 2008, pp. 182--191.
\newblock \href {https://doi.org/10.1109/ICPC.2008.24}
  {\path{doi:10.1109/ICPC.2008.24}}.

\bibitem{cohens:kappa}
J.~Cohen, \href{http://dx.doi.org/10.1177/001316446002000104}{A coefficient of
  agreement for nominal scales}, Educational and Psychological Measurement
  20~(1) (1960) 37--46.
\newblock \href
  {http://arxiv.org/abs/http://dx.doi.org/10.1177/001316446002000104}
  {\path{arXiv:http://dx.doi.org/10.1177/001316446002000104}}, \href
  {https://doi.org/10.1177/001316446002000104}
  {\path{doi:10.1177/001316446002000104}}.
\newline\urlprefix\url{http://dx.doi.org/10.1177/001316446002000104}

\bibitem{charmaz2014:CGT}
K.~Charmaz, Constructing grounded theory, Sage Publishing, London, UK, 2014.

\bibitem{gt:three:elem}
N.~R. Pandit, \href{https://ueaeprints.uea.ac.uk/28591/}{The creation of
  theory: A recent application of the grounded theory method}, The Qualitative
  Report 2~(4) (1996) 1--20.
\newline\urlprefix\url{https://ueaeprints.uea.ac.uk/28591/}

\bibitem{mann:whitney:original}
H.~B. Mann, D.~R. Whitney, \href{http://www.jstor.org/stable/2236101}{On a test
  of whether one of two random variables is stochastically larger than the
  other}, The Annals of Mathematical Statistics 18~(1) (1947) 50--60.
\newline\urlprefix\url{http://www.jstor.org/stable/2236101}

\bibitem{cliff1993:original}
N.~Cliff, {Dominance statistics: Ordinal analyses to answer ordinal
  questions.}, Psychological Bulletin 114~(3) (1993) 494--509.

\bibitem{Romano:CliffsCutoff2006}
J.~Romano, J.~Kromrey, J.~Coraggio, J.~Skowronek, {Appropriate statistics for
  ordinal level data: Should we really be using t-test and Cohen'sd for
  evaluating group differences on the NSSE and other surveys?}, in: annual
  meeting of the Florida Association of Institutional Research, 2006, pp. 1--3.

\bibitem{Cutler:RF:FeatImp2007}
D.~R. Cutler, T.~C. Edwards, K.~H. Beard, A.~Cutler, K.~T. Hess, J.~Gibson,
  J.~J. Lawler, \href{http://dx.doi.org/10.1890/07-0539.1}{Random forests for
  classification in ecology}, Ecology 88~(11) (2007) 2783--2792.
\newblock \href {https://doi.org/10.1890/07-0539.1}
  {\path{doi:10.1890/07-0539.1}}.
\newline\urlprefix\url{http://dx.doi.org/10.1890/07-0539.1}

\bibitem{GENUER2010:FeatImp:RF}
R.~Genuer, J.-M. Poggi, C.~Tuleau-Malot,
  \href{http://www.sciencedirect.com/science/article/pii/S0167865510000954}{Variable
  selection using random forests}, Pattern Recognition Letters 31~(14) (2010)
  2225 -- 2236.
\newblock \href {https://doi.org/https://doi.org/10.1016/j.patrec.2010.03.014}
  {\path{doi:https://doi.org/10.1016/j.patrec.2010.03.014}}.
\newline\urlprefix\url{http://www.sciencedirect.com/science/article/pii/S0167865510000954}

\bibitem{themidpoint:garland}
R.~Garland, The mid-point on a rating scale: Is it desirable, Marketing
  Bulletin (1991) 66--70.

\bibitem{Menzies:TSE2007}
T.~Menzies, J.~Greenwald, A.~Frank, Data mining static code attributes to learn
  defect predictors, IEEE Transactions on Software Engineering 33~(1) (2007)
  2--13.
\newblock \href {https://doi.org/10.1109/TSE.2007.256941}
  {\path{doi:10.1109/TSE.2007.256941}}.

\bibitem{Nagappan:ICSE2006}
N.~Nagappan, T.~Ball, A.~Zeller,
  \href{http://doi.acm.org/10.1145/1134285.1134349}{Mining metrics to predict
  component failures}, in: Proceedings of the 28th International Conference on
  Software Engineering, ICSE '06, ACM, New York, NY, USA, 2006, pp. 452--461.
\newblock \href {https://doi.org/10.1145/1134285.1134349}
  {\path{doi:10.1145/1134285.1134349}}.
\newline\urlprefix\url{http://doi.acm.org/10.1145/1134285.1134349}

\bibitem{Ghotra:FeatSelect:MSR2017}
B.~Ghotra, S.~Mcintosh, A.~E. Hassan,
  \href{https://doi.org/10.1109/MSR.2017.18}{A large-scale study of the impact
  of feature selection techniques on defect classification models}, in:
  Proceedings of the 14th International Conference on Mining Software
  Repositories, MSR '17, IEEE Press, Piscataway, NJ, USA, 2017, pp. 146--157.
\newblock \href {https://doi.org/10.1109/MSR.2017.18}
  {\path{doi:10.1109/MSR.2017.18}}.
\newline\urlprefix\url{https://doi.org/10.1109/MSR.2017.18}

\bibitem{scikit:jml}
F.~Pedregosa, G.~Varoquaux, A.~Gramfort, V.~Michel, B.~Thirion, O.~Grisel,
  M.~Blondel, P.~Prettenhofer, R.~Weiss, V.~Dubourg, J.~Vanderplas, A.~Passos,
  D.~Cournapeau, M.~Brucher, M.~Perrot, E.~Duchesnay,
  \href{http://dl.acm.org/citation.cfm?id=1953048.2078195}{Scikit-learn:
  Machine learning in python}, J. Mach. Learn. Res. 12 (2011) 2825--2830.
\newline\urlprefix\url{http://dl.acm.org/citation.cfm?id=1953048.2078195}

\bibitem{tracy:tse:lrisgood:2012}
T.~Hall, S.~Beecham, D.~Bowes, D.~Gray, S.~Counsell, A systematic literature
  review on fault prediction performance in software engineering, IEEE
  Transactions on Software Engineering 38~(6) (2012) 1276--1304.
\newblock \href {https://doi.org/10.1109/TSE.2011.103}
  {\path{doi:10.1109/TSE.2011.103}}.

\bibitem{Rahman:2013:ProcessBetter}
F.~Rahman, P.~Devanbu,
  \href{http://dl.acm.org/citation.cfm?id=2486788.2486846}{How, and why,
  process metrics are better}, in: Proceedings of the 2013 International
  Conference on Software Engineering, ICSE '13, IEEE Press, Piscataway, NJ,
  USA, 2013, pp. 432--441.
\newline\urlprefix\url{http://dl.acm.org/citation.cfm?id=2486788.2486846}

\bibitem{Ghotra:ICSE2015}
B.~Ghotra, S.~McIntosh, A.~E. Hassan,
  \href{http://dl.acm.org/citation.cfm?id=2818754.2818850}{Revisiting the
  impact of classification techniques on the performance of defect prediction
  models}, in: Proceedings of the 37th International Conference on Software
  Engineering - Volume 1, ICSE '15, IEEE Press, Piscataway, NJ, USA, 2015, pp.
  789--800.
\newline\urlprefix\url{http://dl.acm.org/citation.cfm?id=2818754.2818850}

\bibitem{Lessmann:TSE2008}
S.~Lessmann, B.~Baesens, C.~Mues, S.~Pietsch,
  \href{http://dx.doi.org/10.1109/TSE.2008.35}{Benchmarking classification
  models for software defect prediction: A proposed framework and novel
  findings}, IEEE Trans. Softw. Eng. 34~(4) (2008) 485--496.
\newblock \href {https://doi.org/10.1109/TSE.2008.35}
  {\path{doi:10.1109/TSE.2008.35}}.
\newline\urlprefix\url{http://dx.doi.org/10.1109/TSE.2008.35}

\bibitem{Menzies:TSE2007:PRECISION:INSTABILITY}
T.~Menzies, A.~Dekhtyar, J.~Distefano, J.~Greenwald,
  \href{http://dx.doi.org/10.1109/TSE.2007.70721}{Problems with precision: A
  response to "comments on 'data mining static code attributes to learn defect
  predictors'"}, IEEE Trans. Softw. Eng. 33~(9) (2007) 637--640.
\newblock \href {https://doi.org/10.1109/TSE.2007.70721}
  {\path{doi:10.1109/TSE.2007.70721}}.
\newline\urlprefix\url{http://dx.doi.org/10.1109/TSE.2007.70721}

\bibitem{Chakkrit:TSE2017}
C.~Tantithamthavorn, S.~McIntosh, A.~E. Hassan, K.~Matsumoto,
  \href{https://doi.org/10.1109/TSE.2016.2584050}{An empirical comparison of
  model validation techniques for defect prediction models}, IEEE Trans. Softw.
  Eng. 43~(1) (2017) 1--18.
\newblock \href {https://doi.org/10.1109/TSE.2016.2584050}
  {\path{doi:10.1109/TSE.2016.2584050}}.
\newline\urlprefix\url{https://doi.org/10.1109/TSE.2016.2584050}

\bibitem{walden:issre:vpm:tm}
J.~Walden, J.~Stuckman, R.~Scandariato, Predicting vulnerable components:
  Software metrics vs text mining, in: 2014 IEEE 25th International Symposium
  on Software Reliability Engineering, 2014, pp. 23--33.
\newblock \href {https://doi.org/10.1109/ISSRE.2014.32}
  {\path{doi:10.1109/ISSRE.2014.32}}.

\bibitem{porter:original:1997}
M.~F. Porter, \href{http://dl.acm.org/citation.cfm?id=275537.275705}{Readings
  in information retrieval}, Morgan Kaufmann Publishers Inc., San Francisco,
  CA, USA, 1997, Ch. An Algorithm for Suffix Stripping, pp. 313--316.
\newline\urlprefix\url{http://dl.acm.org/citation.cfm?id=275537.275705}

\bibitem{Landis:Koch:Kappa:Range}
J.~R. Landis, G.~G. Koch, \href{http://www.jstor.org/stable/2529310}{The
  measurement of observer agreement for categorical data}, Biometrics 33~(1)
  (1977) 159--174.
\newline\urlprefix\url{http://www.jstor.org/stable/2529310}

\bibitem{Henningsson:ISESE2004}
K.~Henningsson, C.~Wohlin,
  \href{http://dx.doi.org/10.1109/ISESE.2004.13}{Assuring fault classification
  agreement " an empirical evaluation}, in: Proceedings of the 2004
  International Symposium on Empirical Software Engineering, ISESE '04, IEEE
  Computer Society, Washington, DC, USA, 2004, pp. 95--104.
\newblock \href {https://doi.org/10.1109/ISESE.2004.13}
  {\path{doi:10.1109/ISESE.2004.13}}.
\newline\urlprefix\url{http://dx.doi.org/10.1109/ISESE.2004.13}

\bibitem{Devanbu:Belief:Evidence}
P.~Devanbu, T.~Zimmermann, C.~Bird,
  \href{http://doi.acm.org/10.1145/2884781.2884812}{Belief \&\#38; evidence in
  empirical software engineering}, in: Proceedings of the 38th International
  Conference on Software Engineering, ICSE '16, ACM, New York, NY, USA, 2016,
  pp. 108--119.
\newblock \href {https://doi.org/10.1145/2884781.2884812}
  {\path{doi:10.1145/2884781.2884812}}.
\newline\urlprefix\url{http://doi.acm.org/10.1145/2884781.2884812}

\bibitem{ICSE2013:BugPredictionForHuman}
C.~Lewis, Z.~Lin, C.~Sadowski, X.~Zhu, R.~Ou, E.~J. Whitehead~Jr.,
  \href{http://dl.acm.org/citation.cfm?id=2486788.2486838}{Does bug prediction
  support human developers? findings from a google case study}, in: Proceedings
  of the 2013 International Conference on Software Engineering, ICSE '13, IEEE
  Press, Piscataway, NJ, USA, 2013, pp. 372--381.
\newline\urlprefix\url{http://dl.acm.org/citation.cfm?id=2486788.2486838}

\end{thebibliography}

\end{document}